\documentclass{article}

\usepackage{graphics}
\usepackage{graphicx}
\usepackage{caption}
\usepackage{subcaption}
\usepackage{adjustbox}
\usepackage{amsmath}
\usepackage{amsthm}
\usepackage{amssymb}
\usepackage{amsfonts}
\usepackage{url}
\usepackage{bbold}
\usepackage[utf8]{inputenc}
\usepackage[english]{babel}
\usepackage{float}
\usepackage{listings}
\usepackage[letterpaper, margin=1in]{geometry}
\usepackage{mathtools}
\usepackage{multirow}
\usepackage{natbib}
\usepackage{apalike}
\usepackage{xcolor}
\usepackage{rotating}
\usepackage{enumitem}
\usepackage{lscape}
\usepackage{authblk}

\renewcommand{\thesection}{\arabic{section}}

\makeatletter
\newcommand{\vast}{\bBigg@{3}}
\newcommand{\Vast}{\bBigg@{3}}
\makeatother

\newcommand\independent{\protect\mathpalette{\protect\independenT}{\perp}}
\def\independenT#1#2{\mathrel{\rlap{$#1#2$}\mkern2mu{#1#2}}}

\newcommand{\PM}{PM$_{2.5}$}
\newcommand{\mugm}{$\mu$g/m$^3$}

\theoremstyle{plain}
\newtheorem{assumption}{Assumption}

\newcommand{\beginsupplement}{
    \setcounter{table}{0}
    \renewcommand{\thetable}{S\arabic{table}}%
    \setcounter{figure}{0}
    \renewcommand{\thefigure}{S\arabic{figure}}%
    \setcounter{section}{0}%
    \renewcommand{\thesection}{S\arabic{section}}%
}

\DeclareMathOperator*{\argmin}{arg\,min}

\begin{document}

\title{Estimating a Causal Exposure Response Function with a Continuous Error-Prone Exposure: A Study of Fine Particulate Matter and All-Cause Mortality}

\author[1,*]{Kevin P. Josey}
\author[2]{Priyanka deSouza}
\author[3,4]{Xiao Wu}
\author[1,5]{Danielle Braun}
\author[1]{Rachel Nethery}
\affil[1]{Department of Biostatistics, Harvard T.H. Chan School of Public Health, Boston, MA}
\affil[2]{Department of Urban and Regional Planning, University of Colorado, Denver, CO}
\affil[3]{Department of Statistics, Stanford University, Stanford, CA}
\affil[4]{Stanford Data Science, Stanford University, Stanford, CA}
\affil[5]{Department of Data Science, Dana-Farber Cancer Institute, Boston, MA}
\affil[*]{Corresponding Author: kjosey@hsph.harvard.edu}

\maketitle

\begin{abstract}
Numerous studies have examined the associations between long-term exposure to fine particulate matter (\PM) and adverse health outcomes. Recently, many of these studies have begun to employ high-resolution predicted \PM\ concentrations, which are subject to measurement error. Previous approaches for exposure measurement error correction have either been applied in non-causal settings or have only considered a categorical exposure. Moreover, most procedures have failed to account for uncertainty induced by error correction when fitting an exposure-response function (ERF). To remedy these deficiencies, we develop a multiple imputation framework that combines regression calibration and Bayesian techniques to estimate a causal ERF. We demonstrate how the output of the measurement error correction steps can be seamlessly integrated into a Bayesian additive regression trees (BART) estimator of the causal ERF. We also demonstrate how locally-weighted smoothing of the posterior samples from BART can be used to create a more accurate ERF estimate. Our proposed approach also properly propagates the exposure measurement error uncertainty to yield accurate standard error estimates. We assess the robustness of our proposed approach in an extensive simulation study. We then apply our methodology to estimate the effects of \PM\ on all-cause mortality among Medicare enrollees in New England from 2000-2012.
\end{abstract}

\section{Introduction}

Methods for conducting causal inference with continuous exposures have gained significant traction in recent years. The goal of these methods is to estimate a causal exposure response function (ERF) that is free of confounding bias \citep{kennedy2017nonparametric,wu2018matching}. Estimators of an ERF typically rely on estimates of either the generalized propensity score (GPS), a marginalized model of the outcome process, or some combination of the two to adjust for confounding. However, most of these methods make the implicit assumption that the exposure is measured without error, which is implausible in many observational study settings. For example, in modern environmental health research, air pollution exposure measurements are often derived from model-based predictions of pollutant concentrations rather than the exact pollutant concentrations experienced by an individual. Moreover, because many studies record only areal measures of residential locations such as ZIP-codes, cities, or counties, exposure measures often represent aggregated predictions across these areas. Using aggregated exposures assumes that concentrations are homogeneous within areas and experienced by all individuals residing in the area. As such, exposure measurement error is prevalent in many air pollution epidemiological studies \citep{kioumourtzoglou2014exposure}. Using an error-prone exposure (EPE) in place of the true exposure violates standard causal inference assumptions and may result in biased ERF estimates. Due to the policy-relevance of air pollution ERFs, and particularly those produced using causal inference techniques, invalid inferences caused by EPEs could have severe consequences.

While measurement error has been studied extensively outside of causal inference settings \citep{carroll2006measurement,cole2006multiple}, accounting for EPEs in causal inference is a relatively new endeavor and is mainly confined to scenarios with binary and categorical exposures \citep{lewbel2007estimation, braun2017propensity, wu2019causal}, or cases of measurement error in a confounder instead of the exposure \citep{lenis2017doubly,webb2017imputation}. Beyond the issues encountered when using an EPE in a typical generalized linear model setting, accommodating an EPE in a causal model presents additional challenges in conjunction with resolving confounding bias \citep{braun2017propensity}, many of which remain unaddressed. Moreover, propagating measurement error-related uncertainty into ERF estimates is paramount for proper inference in not just causal settings, but more broadly for many environmental epidemiological studies.

We develop a multiple imputation framework for estimating causal ERFs which incorporates corrections for various types of exposure measurement error commonplace in environmental epidemiology. This approach jointly samples from three sub-models: 1) a model of the EPE, which imputes the true exposures using information from covariates and validation data when available, 2) a GPS model, which improves the accuracy of the imputations, and 3) an outcome model, which estimates the ERF using the imputed exposures while adjusting for confounders. Markov Chain Monte Carlo (MCMC) methods are used to sample from the posterior distributions of the unobserved true exposures and the expected outcome conditioned on the exposures and confounders. A Bayesian additive regression trees (BART) model is specified for the outcome sub-model to capture complex, non-linear relationships between the exposure, the covariates, and the outcome. We also show that implementing a further smoothing step on the BART-estimated ERF using local regression techniques provides accurate estimates of the ERF with fewer posterior samples than a classic Bayesian approach, all while adequately propagating the measurement error uncertainty. The latter smoothing step transitions our approach from a Bayesian estimator into one better characterized as a multiple imputation estimator, while at the same time producing smoother estimates of the ERF than the notoriously noisy BART output \citep{nethery2020evaluation}.

The remainder of this article is structured as follows. We describe the motivating data example in Section~\ref{motivate}. In Section~\ref{prelim} we introduce notation, define the measurement error sources, and identify the  causal ERF model. Section~\ref{bayes} outlines the procedure for correcting the attenuation bias created by measurement error using multiple imputation methods, and we discuss several important caveats and limitations to our proposed methodology. Section~\ref{sim} contains a simulation study. Section~\ref{example} applies the proposed method to create a measurement error-corrected causal ERF for long-term fine particulate matter (\PM) exposure and all-cause mortality in the Medicare population in New England, using the data described in Section~\ref{motivate}. We conclude with a discussion in Section~\ref{discussion}.

\subsection{Motivating Example}\label{motivate}

\PM\ is a well-studied air pollutant known to adversely impact numerous health outcomes \citep{hajat2002effects, dominici2006fine, brook2010particulate, zhu2017short}, including all-cause mortality \citep{anderson2009air,di2017air,wu2020evaluating}, cardiovascular disease \citep{dominici2006fine,zanobetti2009fine,pope2015relationships,yazdi2019long,danesh2021long}, and pulmonary/respiratory diseases \citep{dominici2006fine,zanobetti2009fine,rhee2019impact}. In one of these studies, using a cohort of Medicare enrollees 2000-2012, \cite{wu2019causal} implemented various causal inference techniques to assess whether long-term \PM\ exposure increases the risk of all-cause mortality among older Americans ($>65$ years of age). While \cite{wu2019causal} employed a regression calibration approach to resolve parts of the exposure measurement error present in this application, there were other error components that were left uncorrected. Moreover, a simple regression calibration approach may fail to adequately propagate measurement error-related uncertainty into the causal effect estimates. Additionally, \cite{wu2019causal} considered exposure categories instead of examining the ERF across a continuum of exposures. Categorization of exposures implicitly assumes that participants in the same category are exposed to the same exposure level, which may induce yet another source of measurement error. We seek to investigate the same scientific question as \cite{wu2019causal} using the same data but employing a novel analytic approach that: 1) corrects for additional sources of measurement error in the \PM\ exposures and 2) models the continuous ERF as opposed to categorizing the exposure.

We utilize annual average \PM\ predictions for 1km$\times$1km grid-years across the US from \cite{di2016hybrid}, which were generated via a neural network that combines information from satellite, ground monitor, and land use data. These exposure predictions have been widely used in epidemiological studies \citep{di2017air,rhee2019impact,wu2019causal,yazdi2019long,wu2020evaluating,danesh2021long}. Model assessments indicate that the accuracy of the predictions vary across regions \citep{di2016hybrid}. Because these exposure predictions are error-prone, any inference drawn from them without correction may be biased. To control for this source of measurement error, we use the same ground-monitor \PM\ data that were used to generate the grid-predicted \PM\ measurements in the first place. We treat the measurements from these monitors as the gold standard measurements of \PM\ exposure (i.e. error-free measurements). Our model uses the monitored data as a validation set to re-calibrate the error-prone \PM\ measurements. PM$_{2.5}$ ground monitor locations in New England are overlaid on a ZIP-code map in Figure~\ref{fig:map-locations}. 

Another limitation of the data is that the residential locations of the Medicare enrollees can only be mapped to ZIP-codes \citep{wu2020evaluating}. Thus, following common practice in this setting, we use ZIP-code-years as our units of analysis and all-cause mortality rates within ZIP-code years as our outcome. This creates a misalignment between the \PM\ exposure predictions, which occur on 1km grids, and the ZIP-codes to which we wish to assign \PM\ exposures. 

We also leverage data on demographic and socioeconomic status (SES) factors, as well as land-use variables which are used to aid in measurement error correction. The following potential confounder variables are collected from the US Census Bureau at the ZIP-code-year level: population density, median household income, percent population and households below poverty level, racial/ethnic composition, and distribution of educational attainment (college, some college, high school, not completed high school). Additionally, individual level information about the Medicare enrollees, including their sex, age, race, and Medicaid eligibility status (a proxy for low-income) are obtained from the Medicare record and summarized at the ZIP-code-year level. The land-use variables, collected at the grid-year level, are: surface temperature, accumulated precipitation, radiation flux, accumulated total evaporation, heat flux, precipitation rate, humidity, snow cover, cloud cover, and wind speed. 

\section{Preliminaries}\label{prelim}

\subsection{Notation and Measurement Error}\label{notation}

Our data involve measurements at two different spatial scales: ZIP-code-years (outcome, covariates) and 1km$\times$1km grid-years (exposure, land-use variables). In general, we refer to grid-years as cells, and cells are assumed to be nested within ZIP-code-years, which we refer to as clusters. Let $\mathbf{X}_{i} \in \mathcal{X}$ be a vector of covariates for cluster $i$ ($i = 1,2,\ldots,n$) and $Y_{i}$ denote the outcome variable in that cluster. While the methodology we present easily generalizes to other outcome distributions, in our applied example the $Y_{i}$ are counts (e.g. number of deaths in ZIP-code-year $i$) with $N_i$ representing an offset (e.g. number of at-risk person-years in ZIP-code-year $i$). The combination of these two measures allows us to define the rate $\bar{Y}_i \equiv Y_i/N_i$. We denote the cumulative density functions for the $\mathbf{X}_i$ with $P(\mathbf{x})$ with the associated empirical density function denoted by $\mathbb{P}(\mathbf{x})$.

We assume there is a true exposure both on the cluster level and on the cell level. To enable alignment with the outcome data at the cluster level, the true cluster level exposure, denoted $A_i \in \mathcal{A}$, is the latent variable we are most concerned with accurately predicting. We assume that the $A_i$ are continuous but unobserved for all clusters. The true cell level exposures within cluster $i$ are indexed by $j = 1,2,\ldots,M_i$ and are denoted as $S_{ij}$. The total number of cells is written as $m \equiv \sum_{i = 1}^n M_i$. We assume, as in our example, that the $S_{ij}$ will only be observed for a subset of cells -- the grids containing monitors -- which are indexed by $(i,j) \in \mathcal{S}$. We also have for every cluster and cell an observed model-predicted value of $S_{ij}$, denoted as $\tilde{S}_{ij}$. These predictions are the error-prone exposures. The aggregate value of these EPEs to the cluster level will be denoted with $\tilde{Z}_i = M_i^{-1}\sum_{j = 1}^{M_i} \tilde{S}_{ij}$. Figure~\ref{fig:misalign} illustrates the relationship between $A_i$, $S_{ij}$, and $\tilde{S}_{ij}$ in the context of our motivating application. We also allow for the existence of a set of cell level features, denoted by the vector $\mathbf{W}_{ij}$, that might inform us about the relationship between $S_{ij}$ and $\tilde{S}_{ij}$. In our applied example, the $\mathbf{W}_{ij}$ are land-use variables measured at the grid-year level. $\mathbf{X}_i$ may contain aggregated and/or transformed values of the variables contained in $\mathbf{W}_{ij}$. 

Finally, the conditional expected value of any given variable $D$ is denoted by $\mu_{D}(\cdot) \equiv \mathbb{E}(D|\cdot)$. The conditional probability density function of $D$ is denoted with $p(D|\cdot)$. The generalized propensity score (GPS) is the following particular conditional probability density function: $p(A_i|\mathbf{X}_i)$.

\subsection{Measurement Error Model}\label{measurement}

We formulate the measurement error in this context as a special case of nondifferential classical error. As we noted in Section~\ref{notation}, we must contend with the problem that $S_{ij}$ is rarely observed, and in most areas only the EPE measurements, $\tilde{S}_{ij}$, are available. Moreover, conducting a cluster level analysis relating $A_i$ to $\bar{Y}_i$ requires measurements of $A_i$, not $S_{ij}$. Therefore, we must consider approaches for summarizing $S_{ij}$ while accounting for the possible confounding influence of $\mathbf{X}_i$ and the varying sizes of $M_i$, and ensuring the uncertainty from this process is propagated into the final estimator.

Throughout this paper, we will assume that $\mathbb{E}(S_{ij}|A_i) = A_i$, which can be framed as treating the true cell level exposures $S_{ij}$ as replicate measures of the corresponding cluster level exposure $A_i$, each measured with some amount of error. This `replicate measures' conceptualization is commonly used in the presentation of classical measurement error methods. An additional dimension of complexity is added by the need to rely on cell level exposure predictions, $\tilde{S}_{ij}$. This leads us to formulate the measurement error structure as a variant of the typical nondifferential classical measurement error, which can be decomposed into two components as follows:
\begin{equation}\label{classical}
    U_{ij} = \underbrace{\overbrace{S_{ij} - \tilde{S}_{ij}}^\text{Prediction} + \overbrace{\tilde{S}_{ij} - A_i}^\text{Aggregation}}_\text{Classical}.
\end{equation}
Following conventions in the classical measurement error literature, we make the assumption that the measurement error $U_{ij}$ is homoscedastic and conditionally independent such that:
\begin{assumption}[Conditionally Independent Measurement Error]\label{independence}
We assume $\left(U_{ij} \independent U_{ij'}\right)|A_i, \mathbf{W}_{ij}$ for two cells $j \ne j'$ in a given cluster $i = 1,2,\ldots,n$;
\end{assumption}
\begin{assumption}[Homoscedastic Measurement Error]\label{homoscedasticity}
For all $j = 1,2,\ldots,M_i$ contained within every $i = 1,2,\ldots,n$, we assume the measurement error conditional on $A_i$ has constant variance; $\mathbb{V}(U_{ij}|A_{i}) \equiv \omega^2$.
\end{assumption}

The prediction error component in (\ref{classical}) can be characterized as a Berkson error term \citep{haber2021bias}. The aggregation error, on the other hand, is a type of classical measurement error, implying the composite measurement error term $U_{ij}$ is also a type of classical error. To help cement this understanding, ignore for a moment the prediction error term in $U_{ij}$. The goal of regression calibration is to replace classical measurement error with Berkson error. It has been noted in the regression calibration literature that Berkson error resulting from regression calibration often yields less bias than its classical error counterpart \citep{carroll2006measurement}. Our method, presented in the following sections, is developed in the same spirit, replacing $U_{ij}$ with a Berkson error term via multiple imputation. Our method also allows the user to replace the prediction error component of $U_{ij}$, if desired, but doing so essentially substitutes the original Berkson prediction error with another Berkson error term. That said, re-calibrating the predictions $\tilde{S}_{ij}$ can still be useful if we suspect $\mathbb{E}[\tilde{S}_{ij}|A_i] \ne A_i$. In this latter case, if we assume $\mathbb{E}[S_{ij}|A_i] = A_i$ for all $(i,j) \in \mathcal{S}$, then we can use validation data to find an unbiased predictor for $S_{ij}$, denoted by $\hat{S}_{ij} \equiv \hat{\mu}_S(\tilde{S}_{ij}, W_{ij})$, which implies $\mathbb{E}[\hat{S}_{ij}|A_i] = A_i$.

As we will show in Section~\ref{bayes}, corrections for both the classical measurement error between $S_{ij}$ and $A_i$ and any potential bias found in the predictions $\tilde{S}_{ij}$ can be accommodated under the proposed multiple imputation framework. The imputations of the true exposures are drawn from an MCMC sampler using the full conditional likelihood in (\ref{likelihood}).

\subsection{Assumptions and Identification}\label{assumption}

The methods we employ operate within the Neyman-Rubin causal model modified for continuous exposures \cite{rubin1974estimating}. Let $\bar{Y}_i(a)$ be the outcome that would occur in cluster $i$ if, possibly contrary to what is observed, it had received exposure level $A_i = a$. We refer to the $\bar{Y}_i(a)$ for any $a \in \mathcal{A}$ as the potential outcomes. For some exposure level $a$ the objective analysis is to estimate the marginal ERF, $\theta(a) \equiv \mathbb{E}[\bar{Y}_i(a)]$. Unlike in the binary or categorical exposure setting, where there is only a finite number of unrealized potential outcomes, with a continuous exposure there is an infinite number of unrealized potential outcomes for every unit in the study. Despite this perceived challenge, it is relatively straightforward to translate the assumptions intended for a categorical exposure under the Neyman-Rubin causal model to a continuous exposure setting necessary to identify and estimate $\theta(a)$. 

To start, we invoke the stable unit treatment value assumption which consists of two conditions: 1) consistency and 2) no interference. Consistency refers to the notion that $\bar{Y}_i = \bar{Y}_i(A_i)$, i.e. each unit's observed outcome is equal to its potential outcome evaluated at the true exposure experienced. We are careful to define this assumption in the setup to our problem; recognizing that $A_i$ is unobserved. While unobserved, the true exposure $A_i$ still exists, so consistency should hold. On the other hand, if we na\"ively substitute $\tilde{Z}_i$ for $A_i$, then consistency is likely violated with respect to the EPE unless $\tilde{Z}_i = A_i$ implying $\bar{Y}_i(\tilde{Z}_i) = \bar{Y}_i(A_i)$ for all $i$. As this assumption is unlikely to hold, it is safe to assume consistency is violated if measurement error-agnostic approaches are taken. No interference refers to the assumption that the exposure of one unit does not affect the potential outcomes of another unit. This condition is difficult to uphold in many environmental epidemiology studies, including our own applied example. Major air pollution emissions sources can affect broad areas (e.g., many ZIP-codes), which may lead to instances where the exposure assignment of one ZIP-code affects the outcome of another \citep{papadogeorgou2019causal}. We acknowledge this limitation, yet addressing it is outside the scope of this work.

The fundamental problem in causal inference with any type of exposure is that $\theta(a)$ is not intrinsically estimable since $\bar{Y}_i(a)$ is not observed for every $a \in \mathcal{A}$. However, $\theta(a)$ can still be estimated under the strongly ignorable exposure assumption. With the Neyman-Rubin causal model, this requires the following two assumptions \citep{kennedy2017nonparametric}:
\begin{assumption}[Overlap/positivity]\label{overlap}
$p(A_i|\mathbf{X}_i) > p_{\text{min}} > 0$.
\end{assumption}
\begin{assumption}[Weak unconfoundedness]\label{sita}
$\left[\bar{Y}_i(a) \independent A_i\right] | \mathbf{X}_i$ for all $a \in \mathcal{A}$.
\end{assumption}
\noindent The overlap assumption requires the exposure assignment mechanism to be non-deterministic when conditioned on the covariates. In other words, the probability that a unit is exposed to any level of the exposure along the support is always greater than zero (or $p_{\text{min}}$). Weak unconfoundedness states that the potential outcome evaluated at $a$ does not depend on the true exposure $A_i$ when adjusted by the set of observed confounding variables. This implies there can be no unmeasured confounding. If all confounders for the relationship between $\bar{Y}_i$ and $A_i$ are measured, the latter assumption should hold even when $A_i$ exists but is not observed. When these assumptions hold, it is valid to draw causal conclusions either with experimental or observational data and a continuous exposure \citep{gill2001causal}, and assuming $A_i$ is known, the true causal ERF can be identified as \[ \mathbb{E}\left\{\mathbb{E}\left[\bar{Y}_i\middle|A_i = a, \mathbf{X}_i\right]\right\} = \mathbb{E}\left\{\mathbb{E}\left[\bar{Y}_i(a)\middle|\mathbf{X}_i\right]\right\} = \mathbb{E}\left[\bar{Y}_i(a)\right] = \theta(a) \]

\section{A Multiple Imputation Approach}\label{bayes}

\subsection{Sampling the Exposure and Nuisance Parameters}\label{mcmc}

We introduce our method by first specifying the joint likelihood function for $\bar{Y}_i$, $A_i$, and $S_{ij}$, which incorporates the components needed to address the measurement error described by (\ref{classical}). Supposing that every piece of data were observed, we have:
\begin{equation}\label{likelihood}
\prod_{i = 1}^n \left\{\underbrace{p_{\bar{Y}}(\bar{Y}_i|A_i,\mathbf{X}_i)}_\text{Outcome Model} \times \underbrace{p_{A}(A_i|\mathbf{X}_{i}, \phi_i) \times p_{\phi}(\phi_i|\mathbf{V})}_\text{GPS Model} \times \prod_{j = 1}^{M_i} \underbrace{p_{S}(S_{ij}|A_i)}_\text{EPE Model} \right\}.
\end{equation} 
From a modelling perspective, we assume that the latent exposure variables are conditionally independent and approximately Normal in distribution with $A_i|\mathbf{X}_i, \phi_i \sim \mathcal{N}\left[\mu_A(\mathbf{X}_i, \phi_i), \sigma^2\right]$ and $S_{ij}|A_i \sim \mathcal{N}(A_i, \omega^2)$, in accordance with Assumptions \ref{independence} and \ref{homoscedasticity}. A random effect $\phi_i$ is included in the model for $A_i$ to control for spatial autocorrelation between the cluster level exposures \citep{lee2013carbayes}. We assume $\boldsymbol{\phi} \equiv (\phi_1, \phi_2, \ldots, \phi_n)^T \sim \mathcal{N}\left[\mathbf{0}, \nu^2\mathbf{Q}^{-1}(\mathbf{V}; \rho)\right]$ where $\mathbf{Q}(\mathbf{V}; \rho)$ is a precision matrix controlling the auto-correlation structure as proposed by \cite{leroux2000estimation} given a binary adjacency matrix $\mathbf{V}$ and a hyperparameter $\rho$ controlling the correlation between adjacent clusters. If we assume the $A_i$ are independent, then we can set $\phi_i = 0$ and $p_{\phi}(\phi_i|\mathbf{V}) = 1$ for all $i$. Because $S_{ij}$ is missing, we can substitute $\tilde{S}_{ij}$ for $S_{ij}$ if we suppose $\mathbb{E}[\tilde{S}_{ij}|A_i] = A_i$. However, when given a set of validation data (i.e. a subset where $S_{ij}$ is observed), a more conservative approach is to instead assume $\mathbb{E}[S_{ij}|A_i] = A_i$ and draw a sequence of posterior predictions for $S_{ij}$ for all $j = 1,2,\ldots,M_i$ and $i = 1,2,\ldots,n$. To do this, we can append the model $\prod_{(i,j) \in \mathcal{S}} p_{S}(S_{ij}|\tilde{S}_{ij}, \mathbf{W}_{ij})$ to (\ref{likelihood}) and, supposing that $S_{ij}|\tilde{S}_{ij},\mathbf{W}_{ij} \sim \mathcal{N}\left[ \mu_S(\tilde{S}_{ij},\mathbf{W}_{ij}), \tau^2\right]$, sample values of $S_{ij}$. Note that this addition overspecifies the full-conditional likelihood in (\ref{likelihood}), so doing so is only possible when validation data are available.

We use the above likelihood throughout this manuscript to describe how to sample posterior draws of $A_i$ contemporaneously with posterior predictions for $\mu_{\bar{Y}}(A_i, \mathbf{X}_i)$. We will show in the next section how these posterior samples can be smoothed to create an estimate of the ERF. Relative to the standard regression calibration approach where only a single imputation of $A_i$ is used in fitting the outcome model, using a set of posterior samples (i.e. multiple imputations) of the error-corrected exposures should better propagate uncertainty attributable to the exposure measurement error into the outcome model.

We use a fully Bayesian joint modeling approach for the measurement error, GPS, and outcome models. Parameter values are sampled from the posterior distribution either by Gibbs or Metropolis-Hastings sampling steps, with an added intermediate step to generate posterior predictive samples of the latent variables $A_i$ and, if necessary, $S_{ij}$. We refer to this intermediate step as the imputation stage, whereas the steps involving sampling the parameters are referred to as the analysis stage. See the full details of the sampling algorithm in Supplemental Section S1.

We must carefully consider the form of $\mu_A(\mathbf{X}_i, \phi_i)$ and $\mu_S(\tilde{S}_{ij}, \mathbf{W}_{ij})$ to properly address bias in the EPE model while also accounting for confounding. In our simulation study and data analysis in Sections \ref{sim} and \ref{example}, we assume linear forms for $\mu_A(\mathbf{X}_i, \phi_i) = \phi_i + \mathbf{X}^T_i \boldsymbol{\beta}$ and $\mu_S(\tilde{S}_{ij}, \mathbf{W}_{ij}) = \left(\tilde{S}_{ij}, \mathbf{W}^T_{ij}\right) \boldsymbol{\alpha}$. To better capture nonlinear associations and better avoid issues with model misspecification, splines or Gaussian processes could alternatively be specified \citep{antonelli2020causal, ren2021bayesian}.

If we assume that $\bar{Y}_i$ is a linear function of $A_i$ and $\mathbf{X}_i$, then the parameters determining $p_{\bar{Y}}(\bar{Y}_i|A_i \mathbf{X}_i)$ can be drawn using a Gibbs sampling step, assuming the $N_i$ are fixed and known. However, we prefer to use a data-driven model for $p_{\bar{Y}}(\bar{Y}_i|A_i, \mathbf{X}_i)$ and therefore $\mu_{\bar{Y}}(A_i, \mathbf{X}_i)$, because this model is perhaps the most essential component in (\ref{likelihood}) for finding accurate estimates of $\theta(a)$. To this end, we employ an iteratively updated BART model \citep{chipman2010bart}. Letting $t = 1,2,\ldots,T$ index the trees for a given iteration of the MCMC, the BART model assumes
\begin{equation}\label{bart}
    \bar{Y}_i = \mu_{\bar{Y}}(A_i, \mathbf{X}_i) + \epsilon_i \approx \sum_{t = 1}^{T} g\left(A_i, \mathbf{X}_i; \mathcal{F}_t, \mathcal{G}_t\right) + \epsilon_i.
\end{equation} 
Here, $g(\cdot)$ is a function that bins observations into groups with binary trees formed by the rules contained in $\mathcal{F}_t$, and node means characterized by the set $\mathcal{G}_t$. A BART model differs from other regression tree ensembling methods because of the priors placed on $\mathcal{F}_t$ and $\mathcal{G}_t|\mathcal{F}_t$, which are used to sample values from the approximate posterior distribution of $\mu_{\bar{Y}}(A_i, \mathbf{X}_i)$. Posterior samples are denoted by superscript $(k)$, $k = 1,2,\ldots,K$, e.g., $\mu^{(k)}_{\bar{Y}}(A^{(k)}_i, \mathbf{X}_i) \approx \sum_{t = 1}^{T} g\left(A^{(k)}_i, \mathbf{X}_i; \mathcal{F}^{(k)}_t, \mathcal{G}^{(k)}_t\right)$. In each iteration of our MCMC sampler, the posterior samples of the BART parameters are drawn conditional on the current posterior sample of the exposures $A_i^{(k)}$. The error term is assigned the distribution $\epsilon_i \sim \mathcal{N}(0,\psi^2/N_i)$. 

For model-fitting, we assign conjugate inverse-gamma prior distributions to the variance parameters, $\omega^2, \tau^2, \sigma^2, \nu^2, \psi^2 \sim \mathcal{IG}(0.001, 0.001)$, and set $\rho \sim \mathcal{U}(0,1)$. The regression coefficients for $\mu_S(\cdot)$, $\mu_A(\cdot)$, are each assigned independent Gaussian priors with default values $\boldsymbol{\alpha}, \boldsymbol{\beta} \sim \mathcal{N}(\mathbf{0}, 10^6\mathbf{I})$ where $\mathbf{0}$ is a vector with every entry equal to zero and $\mathbf{I}$ is an identity matrix (with appropriate dimensions). Additional sampling details are provided in Supplemental Section S1. For an example of how to implement a generalized linear model for $\bar{Y}_i|A^{(k)}_i, \mathbf{X}_i$ instead of the BART implementation in (\ref{bart}), please refer to the simulation experiment in Supplemental Section S4.

\subsection{Smoothing BART Output}\label{loess}

While predictions from the BART model (e.g., means of posterior predictive samples) could be used directly to form an estimate of the ERF, these samples do not typically form a smooth function of $\theta(a)$ over $a\in\mathcal{A}$. Smooth ERFs are believed to be most biologically plausible relationship in many epidemiological and biomedical applications, including the effect of \PM\ on all-cause mortality, and are more desirable for identifying causal relationships \citep{kim2018identification}. To resolve this problem, we will project a multiply-imputed pseudo-outcome derived from the BART output onto the support of $\mathcal{A}$ with local regression techniques. We only need a small number of imputations, i.e., a small, thinned subset of the BART posterior samples (indexed by $l = 1,2,\ldots,L$, $L \ll K$), to get smooth, reliable estimates of the ERF. First, we create multiple imputations of the pseudo-outcome as
\begin{equation}\label{pseudo}
    \xi^{(l)}\left(A^{(l)}_i, \mathbf{X}_i, \bar{Y}_i\right) = \left[\bar{Y}_i - \mu^{(l)}_{\bar{Y}}\left(A^{(l)}_i, \mathbf{X}_i\right)\right] + \int_{\mathcal{X}} \mu^{(l)}_{\bar{Y}}\left(A^{(l)}_i, \mathbf{x}\right) d\mathbb{P}(\mathbf{x}).
\end{equation} 
There are two components in this pseudo-outcome. The integral on the right hand side of the addition symbol is the marginal estimate of the ERF at $A^{(l)}_i$. Under the more typical Bayesian framework, we could find $\int_{\mathcal{X}} \mu^{(k)}_{\bar{Y}}\left(a, \mathbf{x}\right) d\mathbb{P}(\mathbf{x})$ for each $k = 1,2,\ldots,K$. This would be equivalent to a Bayesian version of a G-computation estimator for the ERF \citep{keil2018bayesian}. However, this process can be time consuming when iterating for all $K$ posterior samples instead of the $L$ imputations, in addition to yielding a non-smooth ERF. The term on the left-hand side of the addition symbol in (\ref{pseudo}) is the residual error conditioned on $A^{(l)}_i$ and $\mathbf{X}_i$. Since multiple imputation couples Bayesian methodology with regression techniques, it is necessary for each imputation that we approximate the variance for estimates of $\theta(a)$ at each point $a \in \mathcal{A}$ which is facilitated by the addition of this residual error term. Without the residual error component, the variance estimates within each imputation would be incorrect \citep{antonelli2020causal}.

Local regression methods offer perhaps the most flexible means of projecting the pseudo-outcomes in (\ref{pseudo}) onto the support $a \in \mathcal{A}$. Given a bandwidth $h > 0$, the ERF estimates we obtain for each imputation $l = 1,2,\ldots, L$ are denoted by $\hat{\theta}^{(l)}_{h}(a)$. A pointwise estimate of the ERF is summarized by $\bar{\theta}_h(a) = L^{-1}\sum_{l = 1}^L \hat{\theta}^{(l)}_h(a)$. Because we regress each of the imputed pseudo-outcomes onto the support of the exposure, $a \in \mathcal{A}$, the values $\hat{\theta}^{(l)}_{h}(a)$ do not form a proper posterior of $\theta(a)$. This result was noted also by \cite{antonelli2020causal} who found that the posterior distribution of a regression based estimator like $\hat{\theta}^{(l)}_h(a)$ alone did not adequately characterize the uncertainty. To correct for this, our approach to estimation combines a kernel-weighted least-squares regression estimator similar to \cite{kennedy2017nonparametric} with the Bayesian approach of \cite{antonelli2020causal}. Instead of using a bootstrap estimator to find the empirical MCMC standard error used by \cite{antonelli2020causal}, we use an asymptotic standard error estimator derived by \cite{kennedy2017nonparametric}. Details for estimating $\hat{\theta}^{(l)}_{h}(a)$ and the accompanying standard errors for $\bar{\theta}_h(a)$ using multiple imputation combining rules \citep{rubin2004multiple} are provided in Supplemental Section S2 along with the details on how to smooth the BART ERF estimates using kernel-weighted least-squares regression.

A simple regression calibration variation to the above approach would be to find a single imputation (i.e. $L = 1$) of $A_i$, let's say $\hat{A}_i$, then construct an estimator of the outcome mean $\hat{\mu}_{\bar{Y}}(\hat{A}_i, \mathbf{X}_i)$ and substitute that estimator into (\ref{pseudo}) replacing $\mu_{\bar{Y}}^{(1)}\left(A_i^{(1)}, \mathbf{X}_i\right)$. In this single imputation case, the ERF estimator would equal $\hat{\theta}^{(1)}_h(a)$. We will see in the simulation that for a single imputation, the choice of $\hat{A}_i$ is not so straightforward in the presence of measurement error, nor does such an approach adequately propagate the uncertainty created by measurement error without further correction. For more details on the local regression methods we apply, see Supplemental Section S2.

\subsection{Issues Stemming from Congeniality}\label{congenial}

Our model contains two components, an imputation stage and an analysis stage. The imputation stage generates imputations of the latent exposures, $A_i$ and $S_{ij}$, while the analysis stage draws samples from the posterior distributions of the model parameters. Following the multiple imputation literature, the imputation component of our model must condition on the outcome to satisfy assumptions associated with congeniality, explained below. Congeniality states that the imputation and analysis stage models need to utilize the same data \citep{meng1994multiple}, and violations of congeniality can bias parameter estimates. From a Bayesian perspective, non-congeniality can arise due to cutting feedback or modularizing \citep{zigler2013model} the components of a joint model. The limiting distribution from an MCMC of $A^{(l)}_i$, and by extension the limiting distribution of the exposure effect, is ill-defined when cutting feedback between the outcome and exposure model in a Bayesian setting \citep{plummer2015cuts}. Since we are necessarily using the imputations of $A_i$ to fit the analysis model $\mu_{\bar{Y}}(A_i, \mathbf{X}_i)$ and construct an estimate of the ERF, then to satisfy congeniality and avoid biasing the ERF estimate, the imputation models must be conditional on the outcome. This is accomplished in our method by using a fully Bayesian joint model-fitting scheme for the EPE, GPS, and outcome models.

However, sampling the latent exposures $A_i$ conditioned on the outcome creates bi-directional feedback between the traditional ``design stage'' (GPS modeling) and analysis stage that are kept separate in most causal analyses. This seemingly defies research that emphasizes cutting feedback between the GPS and outcome model in a Bayesian causal analysis \citep{zigler2013model}. Notice that we include a model for the GPS in (\ref{likelihood}) that we fit in Section~\ref{mcmc}, yet we do not utilize the GPS in our outcome model nor our estimator of the ERF in Section~\ref{loess}. While the outcome data does not appear in the full conditionals for the GPS model parameters, it is used to generate new predictions of $A_i$ which may indirectly create feedback affecting the GPS model estimates. To counteract the feedback problems created by congeniality, we removed the GPS adjustments from the doubly-robust pseudo-outcome supposed by \cite{kennedy2017nonparametric} (which appears in the Supplemental Section S4). These GPS adjustments appear to provide no utility in our context as demonstrated by a small simulation study contained within Supplemental Section S4. In this simulation, we draw particular attention to cases where the outcome model is misspecified but the GPS is correctly specified. In this scenario, there is almost no difference between using the GPS adjusted pseudo-outcome supported by \cite{kennedy2017nonparametric}, and the pseudo-outcome we suppose in (\ref{pseudo}). We contend that this null result is attributable to the feedback created by the need to use congenial models for the imputation and analysis stages of the MCMC sampler.

\section{Numerical Example}\label{sim}

\subsection{Simulation Design}\label{design}

In this simulation study, we examine the performance of the method described in Section~\ref{bayes} in the presence and absence of exposure measurement error. In addition, we will evaluate the effects of model misspecification across the three components of the likelihood in (\ref{likelihood}), all while examining whether uncertainty is properly propagated into the final ERF estimate.

We test four different methods for measurement error correction. The first method na\"ively ignores any measurement error and assumes $\tilde{Z}_i$ is the true cluster level exposure. The second method examines an extension to the regression calibration approach proposed by \cite{wu2019causal}, which we adapted to consider a continuous exposure, that corrects for prediction error but disregards the remaining classical error in (\ref{classical}) created by the remaining aggregation error. After finding predictions $\hat{S}_{ij}$ for $S_{ij}$ using least-squares regression conditioned on $\tilde{S}_{ij}$ and $\mathbf{W}_{ij}$, a single imputation of $A_i$ is produced by computing $\hat{Z}_i = M_i^{-1}\sum_{j = 1}^{M_i} \hat{S}_{ij}$. We will refer to these two implementations that use either $\tilde{Z}_i$ or $\hat{Z}_i$ as the ``single imputation'' approaches. The third and fourth methods that we examine are two ``multiple imputation" approaches that follow the proposal in Section~\ref{bayes}, with different outcome model specifications. In one of the multiple imputation variants, we let $\mu^{(l)}_{\bar{Y}}\left(A_i^{(l)}, \mathbf{X}_i\right)$ be a BART model. In another, we let $\mu^{(l)}_{\bar{Y}}\left(A_i^{(l)}, \mathbf{X}_i\right)$ be a correctly-specified log-linear Poisson model, assuming unknown coefficient values (see Supplemental Sections S3 and S4 for details). To highlight the impacts of correcting measurement error and propagating uncertainty, we utilize estimation approaches that are as similar as possible across the four competing methods aside from how they correct for measurement error. For the single imputation approaches, we construct the pseudo-outcome in (\ref{pseudo}) in the exact same manner as in the multiple imputation cases, but for a single iteration, i.e. setting $L = 1$. For the single imputation methods, this means using the $\tilde{Z}_i$ or $\hat{Z}_i$ in place of $A^{(l)}_{i}$ and either $\tilde{\mu}_{\bar{Y}}(\cdot)$ or $\hat{\mu}_{\bar{Y}}(\cdot)$ in place of $\mu^{(l)}_{\bar{Y}}(\cdot)$, respectively. We also assume that $\phi_i = 0$ for all $i = 1,2,\ldots,n$ (i.e. $A_i$ is independent), both in the data generation procedure and in the models we fit.

In Supplemental Section S3 we describe the various data generating mechanisms used to construct the simulation scenarios. For each scenario, we ran $500$ iterations, in each one applying each of the four methods described above. In each iteration, we obtain pointwise estimates of $\theta(a)$ at $201$ equally spaced exposure values across the range $a \in [6, 14]$. We report the relative bias, the residual mean square error, and coverage probabilities averaged across all the $500$ iterations and all the $201$ pointwise ERF estimates. We also report these same measurements for predictions at a single exposure level, $a = 11$, averaged across the $500$ iterations. Briefly, we vary $n \in \{400, 800\}$, $m \in \{5n, 10n\}$, $\tau^2 \in \{0,1,2\}$, and $\omega^2 \in \{0,1,2\}$. Note that when $\tau^2 = 0$ the prediction error is absent, whereas when $\omega^2 = 0$ the remaining classical error is absent and $U_{ij}$ is limited to the prediction error. We also vary the degree of misspecification in the component models contained in (\ref{likelihood}) \citep{kang2007demystifying}.

\subsection{Simulation Results}

The main results of this simulation illustrate how adding measurement error can influence estimates of the ERF. Figure~\ref{fig:sim-erf} shows the average ERF estimated with each of the four methods in Section~\ref{design} (relative to the true ERF) under simulation scenarios with no measurement error, prediction error only, classical error only, and both sources of error present. These results demonstrate that generating unbiased imputations of the exposure of interest is critical, whether using a single or multiple imputations. Note that from the description of the scenarios, we have $\mathbb{E}[\tilde{S}_{ij}|A_i] = A_i$ when the EPE model is correctly specified. Therefore the na\"ive approach and the regression calibration approach, which corrects for prediction error, produce similar results as shown by the root mean squared errors displayed in Figure~\ref{fig:sim-erf}. This is because the prediction error is Berkson, so using regression calibration has a negligible effect. Also, note that the regression calibration and multiple imputation approaches perform similarly in scenarios without classical error (Figure~\ref{fig:sim-erf}, top row), but the performance of regression calibration deteriorates when we simulate measurement error with aggregation error present. Meanwhile, the multiple imputation approaches retain accuracy (Figure~\ref{fig:sim-erf}, bottom row). Some bias still lingers in each method, particularly at the extremes of the exposure support (i.e. $a = 6$ and $a = 14$). This is typical of local regression methods where there is an implicit bias-variance tradeoff through the choice of $h$ (which is set to $h = 0.2$ when $n = 800$ and $h = 0.4$ when $n = 400$) \citep{hastie01statisticallearning}. Unsurprisingly, the accuracy of the different methods we test improves as both $n$ and $m$ increase (Table \ref{tab:sim-rslt}).

We can also see in Table \ref{tab:sim-rslt} that the coverage probabilities from the 95\% confidence interval estimates among the multiple imputation approaches offer a marked improvement over the alternatives. When using a correctly specified log-linear outcome model within a multiple imputation setting, we achieve coverage probabilities that match the nominal 95\% confidence level. However, the coverage probabilities are imperfect when using a BART outcome model. BARTs' tendency to under-report the dispersion of point estimates is well-documented \citep{wendling2018comparing, hahn2020bayesian, nethery2020evaluation}. The BART output also seems to produce biased estimates at several points along the curve, contributing to the low coverage probabilities. Despite this limitation, BART remains the preferred outcome model due to its strong predictive abilities even when the correct model specification is unknown (see the following paragraph). Since we average pointwise evaluations over the support $\mathcal{A}$, we can also see how the coverage probabilities are affected by using local regression techniques. This can be seen by examining the single pointwise evaluations at $a = 11$, which are closer to the nominal 95\% confidence level than the average coverage probability over the $201$ pointwise evaluations.

In the second part to our simulation study, where we induce model misspecification, we observe in Table \ref{tab:sim-rslt} that the ERF estimates are largely unaffected by outcome model misspecification alone with a BART model -- remaining relatively unbiased as long as the GPS is still correctly specified. The same result is obviously not true for a parametric log-linear outcome model. Even though BART is supplied the untransformed covariates when model misspecification is present, it is quite adept at identifying complex nonlinear and interacting effects, making outcome model misspecification a moot consideration, so long as Assumptions \ref{overlap} and \ref{sita} hold. That said, the coverage probability decreases considerably in scenarios with a misspecified outcome model relative to the results of the scenario where all models are correctly specified.

When the GPS model is misspecified, the level of bias increases substantially in each of the ERF estimates, both from the single imputation and multiple imputation models. A perplexing exception to this result is when both the GPS and the outcome model are misspecified, in which case the bias returns to the nominal levels observed in cases when the GPS is correctly specified and a BART outcome model is used. Given these results, and the results of the simulation in Supplemental Section S4, it is evident that finding an approximately correct GPS model may be even more important than finding a correct outcome model as we originally suggested.  Doing so might decrease the precision of the imputations for $A_i$. Finally, when the EPE model is ``misspecified", meaning $\mathbb{E}(\tilde{S}_{ij}|A_i) \ne A_i$ implying $\mathbb{E}(\tilde{Z}_{i}|A_i) \ne A_i$, then we see the most severe levels of bias using the na\"ive (no correction) approach. Since the multiple imputation and regression calibration approaches correctly model $S^{(k)}_{ij}$ and $\hat{S}_{ij}$, respectively, such that $\mathbb{E}(S^{(k)}_{ij}|A_i) = A_i$ and $\mathbb{E}(\hat{S}_{ij}|A_i) = A_i$, the prediction bias does not affect the estimates of the exposure response curves using these two approaches.

\section{Applied Example}\label{example}

Recall the example described in Section~\ref{motivate}. \cite{wu2019causal} used regression calibration methods on the same data to correct for measurement error for the grid-year \PM\ predictions. Subsequently, they aggregated the corrected grid-year predictions to the ZIP-code-year level, categorized the imputed exposures based on standard EPA cutoffs, and examined a variety of GPS-based implementations including matching, sub-classification, and weighting to estimate the effect of \PM\ on mortality. Their results suggest that increasing \PM\ exposure led to excess mortality events.

\begin{figure}[H]
	\centering
	\includegraphics[scale = 0.7]{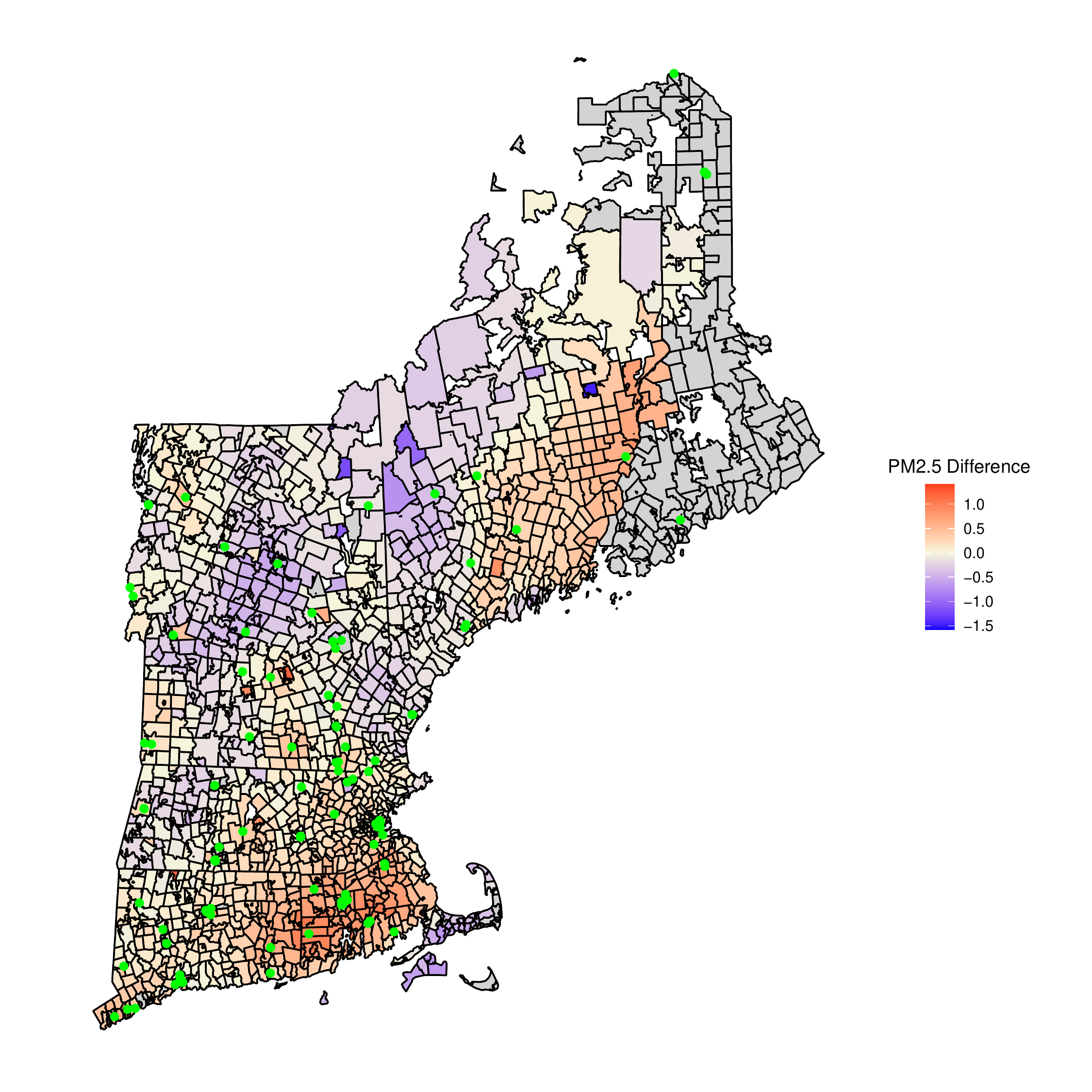}
	\caption{ZIP-code and monitor locations (green dots) appearing across New-England in 2010. The heat map describes the difference between the posterior mean $L^{-1}\sum_{l = 1}^L A^{(l)}_{i}$ and $\tilde{Z}_i$ in 2010. ZIP-codes with a grey fill had either missing exposure data or missing outcome data in 2010.}
	\label{fig:map-locations}
\end{figure}

\begin{figure}[H]
	\centering
	\includegraphics[scale = 0.5]{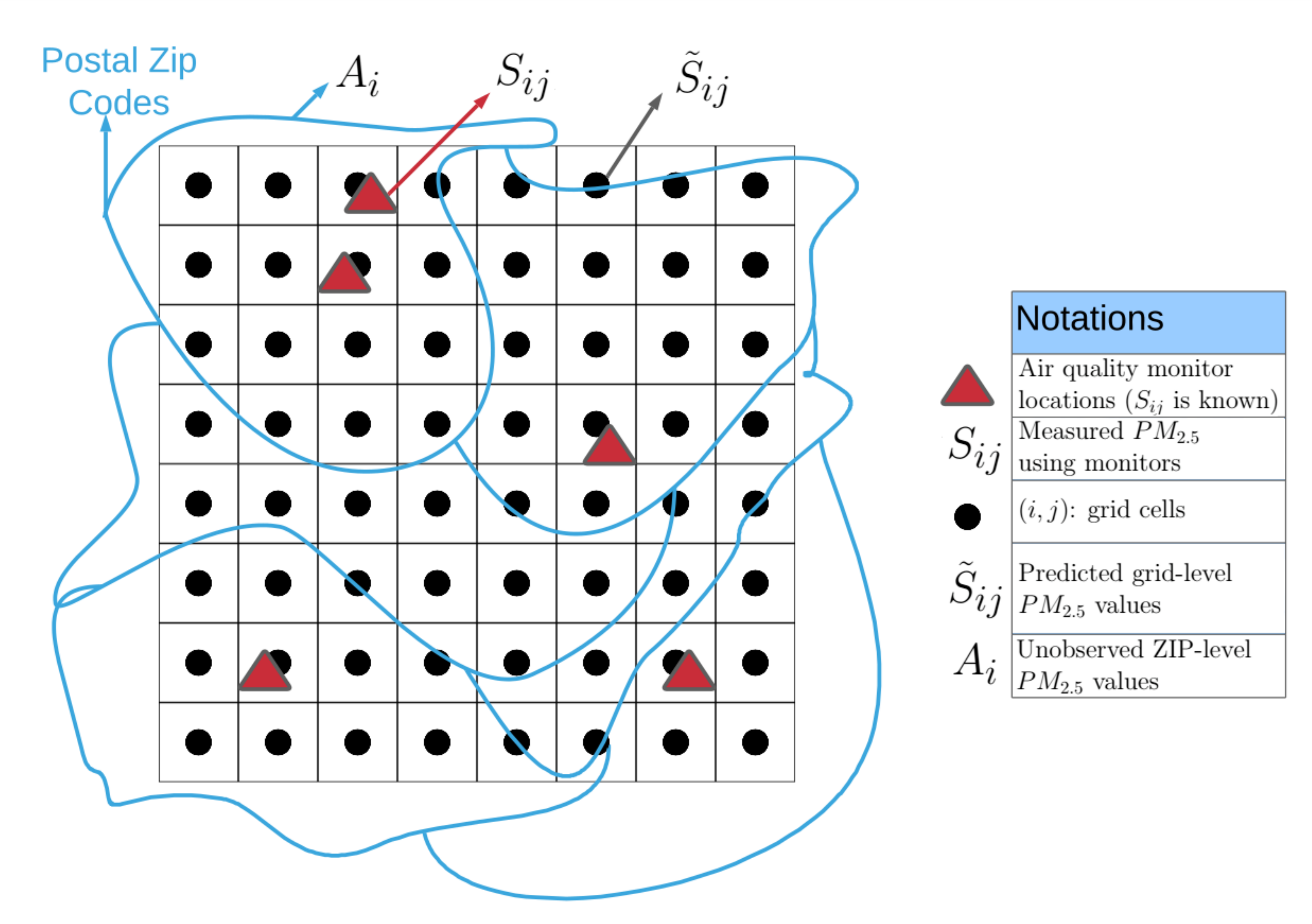}
	\caption{An illustration about the nested structure of the data accompanying the relationship between the true ZIP-code, the true grid level, and the EPE grid measurements.}
	\label{fig:misalign}
\end{figure}

\begin{figure}[H]
	\centering
	\includegraphics[scale = 0.7]{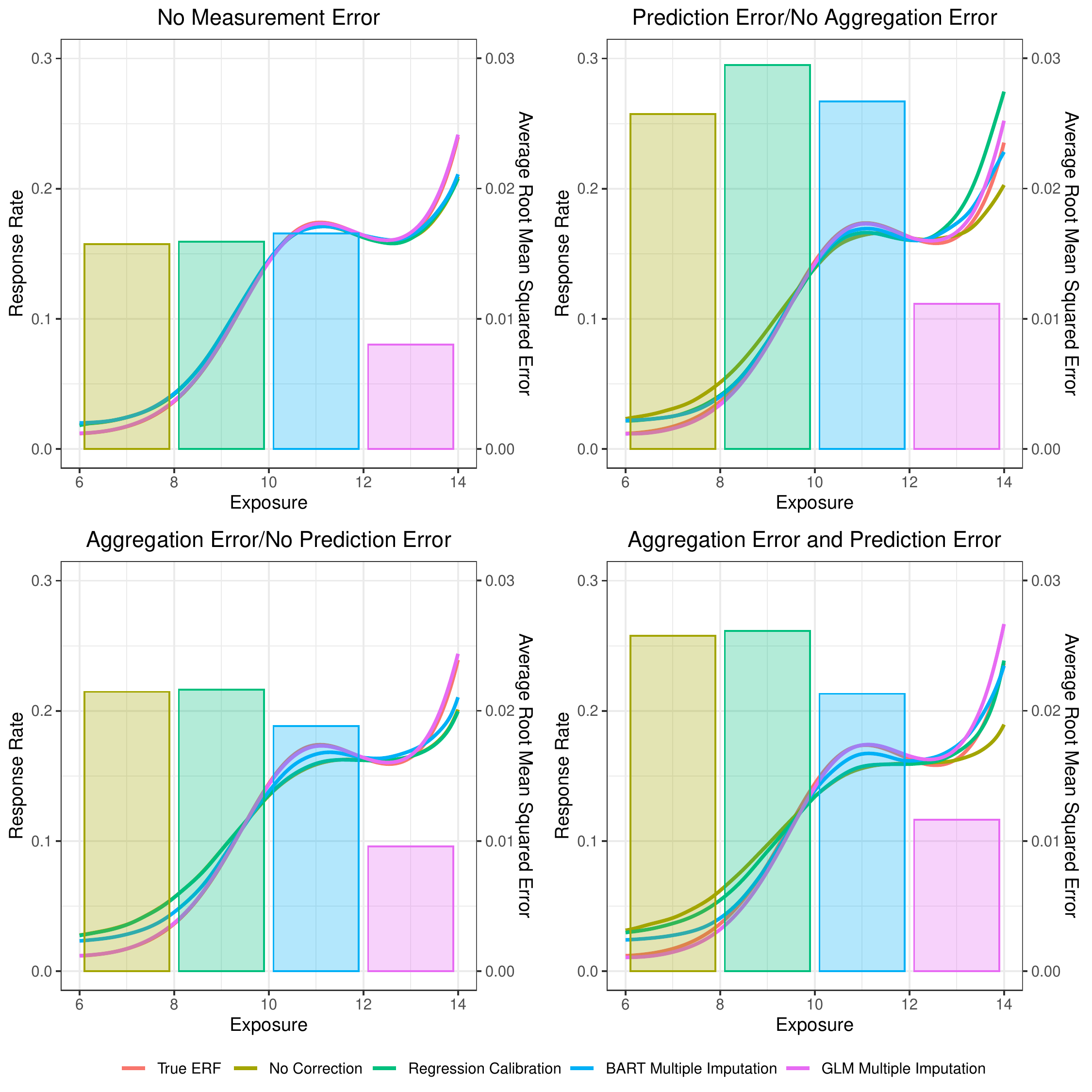}
	\caption{ERF estimates averaged over $500$ simulated iterations when $n = 800$ and $M = 4000$ and correct model specifications. When classical error is present then $\omega^2 = 2$, when prediction error is present then $\tau^2 = 1$. The BART approach to multiple imputation uses a BART outcome model while the GLM implementation uses a log-linear outcome model. The bar plot overlaying the ERF estimates represents the root mean squared errors for the corresponding methods.}\label{fig:sim-erf}
\end{figure}

In this analysis we re-examine the same data as \cite{wu2019causal}, applying the proposed multiple imputation approach to estimate the ERF relating \PM\ with all-cause mortality. For ZIP-code-year $i$, the exposure $A_i$ is the annual average \PM\ concentration (in \mugm), the outcome $Y_i$ is the number of deaths observed, and $N_i$ is the person-years at risk. We have exposure data on $n = 28{,}626$ ZIP-code-years in New England covered by $m = 2{,}395{,}588$ 1km$\times$1km grid-years, each with error-prone \PM\ exposure predictions $\tilde{S}_{ij}$. A subset of $83$ grids within $75$ ZIP-codes have error-free \PM\ measurements from monitoring stations in years 2000-2012 (Figure~\ref{fig:map-locations}). While it is plausible that the $\tilde{S}_{ij}$ and $S_{ij}$ within ZIP-code-years are correlated, the majority of this correlation can be accounted for simply by conditioning on $A_i$, thus enabling Assumptions \ref{independence} and \ref{homoscedasticity}.

For the multiple imputation approach, we assume the outcome model in (\ref{bart}). As in the simulation study, we implement two different single imputation approaches. In the first approach, we ignore prediction error altogether and suppose $\tilde{Z}_i$ is the true exposure (we call this the ``no correction'' approach). In the second approach, we use least squares regression on the validation data to get the regression calibrated $\hat{Z}_i$. We then use these single imputations of the exposure in a BART model to construct an estimate of the mean function; either $\tilde{\mu}_{\bar{Y}}(\tilde{Z}_i, \mathbf{X}_i)$ or $\hat{\mu}_{\bar{Y}}(\hat{Z}_i, \mathbf{X}_i)$. For both the single ($L = 1$) and multiple imputation approaches ($L = 100$), local regression was applied according to Section~\ref{loess}. We collected $5{,}000$ MCMC samples, thinned by every $50$ iterations, after a burn-in of another $5{,}000$ iterations. Diffuse priors specified in Section~\ref{mcmc} were used for the GPS model parameters, the EPE model parameters, as well as for the BART model. As opposed to the simulation study, we account for spatial correlation between the ZIP-code exposures by sampling $\phi_i$ in (\ref{likelihood}). In this example, the matrix $\mathbf{V}$ is a binary adjacency matrix with each row and column representing a different ZIP-code-year.

\begin{figure}[H]
	\centering
	\includegraphics[scale = 0.75]{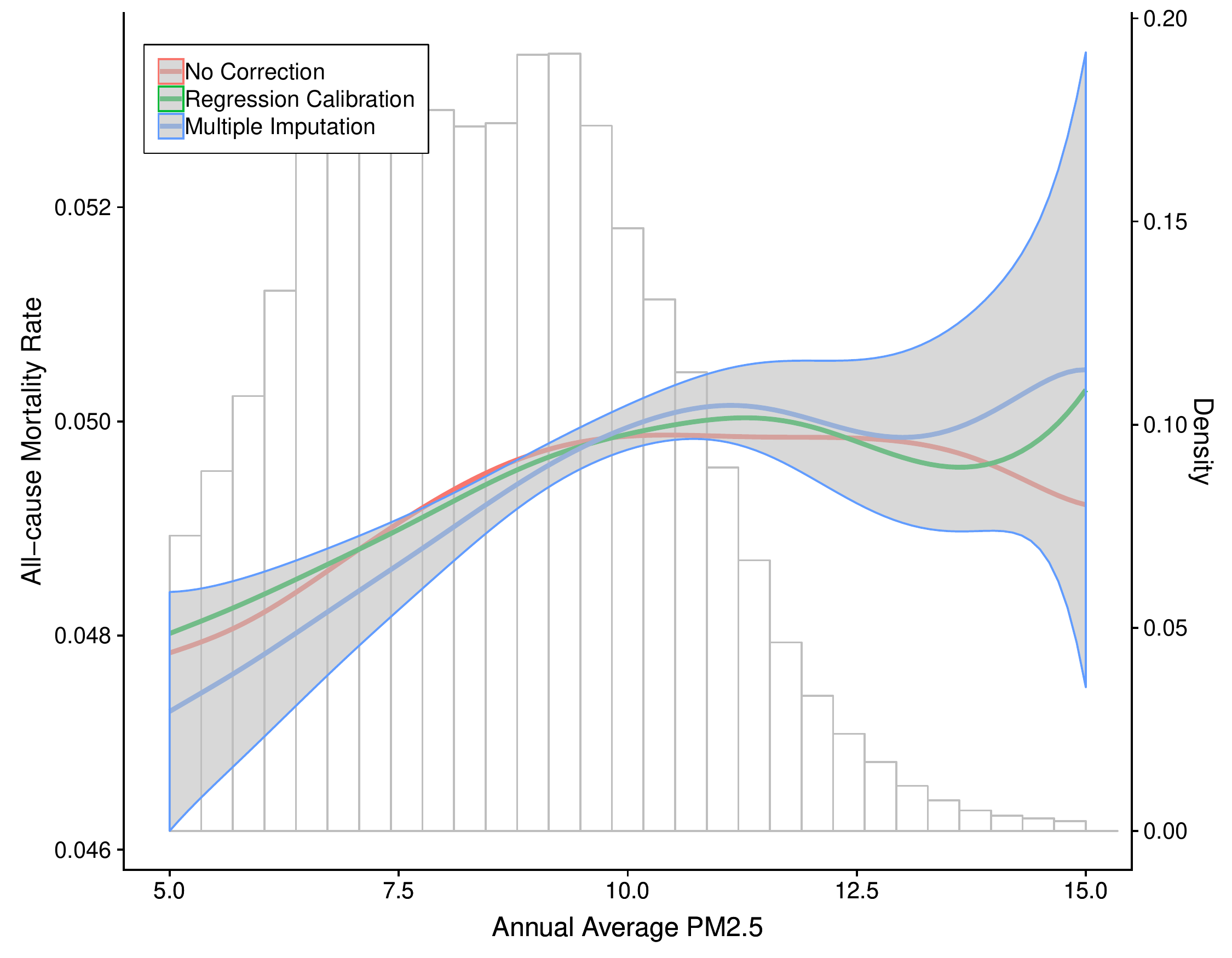}
	\caption{ERF estimate of \PM\ on all-cause mortality in New England between 2000-2012 amongst Medicare recipients under three different approaches to measurement error correction. The grey ribbon represents the 95\% confidence interval computed from our multiple imputation approach. The histogram underlying the curves corresponds with the empirical distribution of the EPEs.} \label{fig:ne-erf}
\end{figure}

Figure~\ref{fig:map-locations} provides a heat-map displaying the difference between the na\"ive exposure imputations of $\tilde{Z}_i$ and the posterior means $L^{-1}\sum_{l = 1}^L A^{(l)}_{i}$ for each ZIP-code in 2010. Here we can see substantial differences in the imputed exposure values in some areas of New England. Figure \ref{fig:ne-erf} shows the ERF estimates from each of the three methods we examined. In the curve estimated with multiple imputation, we can see a modest increase in the rate of all-cause mortality from $4.7\%$ when annual average \PM\ is at $5$ \mugm\ to $5.0\%$ when annual average \PM\ is at $15$ \mugm. Relative to the uncorrected and regression calibration curves, the curve estimated with multiple imputation has a steeper gradient at lower levels of PM$_{2.5}$. The steepest increase in mortality occurs when \PM\ changes from about $5$ \mugm to $10$ \mugm, which is policy-relevant given the World Health Organization's recent decision to lower the recommended limit of annual average \PM\ to $5$ \mugm\ \citep{who2021}.

We suspect that the minor differences between the three curves are primarily attributable to the small variation of the classical measurement error, $\omega^2$, which we found to be equal to about 0.0187 ($95\%$ CI: 0.0183-0.0189). The posterior mean for the conditional prediction error variance, $\tau^2$, on the other hand was 1.340 ($95\%$ CI: 1.212, 1.446). However, as we demonstrated in the simulation study, correcting for the prediction error is secondary to correcting for the classical measurement error including the aggregation error, except when $\mathbb{E}(\tilde{S}_{ij}|A_i) \ne A_i$. This reasoning is further cemented in this illustrative example -- we can see that $\mathbb{E}(\tilde{S}_{ij}|A_i) \approx A_i$, and so the difference between the curves with no measurement error correction and regression calibration is relatively small.

\section{Discussion}\label{discussion}

We developed a multiple imputation framework that addresses exposure measurement error when estimating a causal ERF. We adapted a Bayesian approach for correcting measurement error to generate imputations of the true exposures assuming non-differential classical measurement error. We then constructed a flexible representation of the ERF using kernel smoothed output from a Bayesian additive regression tree. These two steps interface with one another over several iterations. Since the final estimator isn't formally Bayesian due to post-processing the BART adjusted pseudo-outcome with kernel weighted least squares regression, we regard it as a multiple imputation-based estimator. We then described some issues that we encountered with this approach regarding the seeming conflict that exists between cutting feedback and appeasing rules about congeniality. We provided several simulated numerical examples that showcase why correcting for exposure measurement error within a causal-analysis is important to obtain unbiased estimates and valid inferences.

In the real data example, we estimated the exposure response function associating all-cause mortality with annual average \PM\ exposures. At first glance, the three curves we estimate with varying degrees of measurement error correction all seem fairly similar for this particular analysis. These results seem to indicate that there is little benefit to correcting for measurement error while estimating ERFs, though the attenuation at lower \PM\ levels is meaningful. However, the reason for the similarities are mainly due to the low levels of measurement error. As measurement error increases, so does the attenuation bias. Indeed, our attempts to correct for measurement error in our applied example does not seem to have a large impact on the results associating \PM\ with mortality. Therefore, this analysis helps validate other analyses that used similar data but did not correct for measurement error.

While Medicare data limitations necessitate use of ZIP-code aggregate exposures in our analysis, there are various simplifications to Sections \ref{prelim} and \ref{bayes} that can be made to accommodate scenarios where certain measurement error sources are absent.  If we were provided individual level outcomes and error-prone exposure data, then the measurement error would be closer to a Berkson-type rather than classical \citep{bateson2010regression}. Accounting for measurement error in this problem would require a different EPE model than the one proposed, however, adapting our framework should be straightforward given the adaptability of the Bayesian framework that we outlined \citep{carroll2006measurement}. Lastly, our approach assumes non-differential measurement error, which in our context means that $Y_i$ is unaffected by $\tilde{S}_{ij}$ given $A_i$. More work is required to generalize this methodology to be useful in cases where the measurement error is differential.

Another line of future work involves relaxing the independence and homoscedastic assumptions surrounding the measurement error (Assumptions \ref{independence} and \ref{homoscedasticity}). Since these models are deployed within an MCMC framework, more sophisticated spatiotemporal modelling techniques might be employed to better understand the measurement error residing in the data. Second, alternatives to the parametric GPS and EPE models were not implemented in this work. However, more flexible nonparametric methods may help reduce model-dependency of the ERF. We relaxed some of this model-dependency by using a BART outcome model. However, the BART implementation we employed assumes the outcome is approximately Gaussian. There is an option for fitting log-linear BART models \citep{murray2021log} which might ameliorate some of the problems we encountered with BART, like the poor coverage probabilities observed in the simulation study \citep{hahn2020bayesian}. However, at the time of writing, code to fit this alternative BART model could not be obtained. This problem may also be addressed by choosing a different outcome model as demonstrated in the simulations.

\section*{Acknowledgements}

Funding was provided by the National Institute of Health (NIH) grants (T32 ES007142, R01 ES030616,  R01 ES028033, R01 ES026217, K01 ES032458-01) and HEI grant 4953-RFA14-3/16-4. The contents are solely the responsibility of the grantee and do not necessarily represent the official views of the funding agencies. Further, the funding agencies do not endorse the purchase of any commercial products or services related to this publication.

\section*{Reproducible R Code}

The \texttt{R} functions used to fit the model in Section~\ref{bayes} along with the code used to run the simulation studies in Sections \ref{sim} and S4 are available at \texttt{https://github.com/kevjosey/causal-me/}.

\bibliographystyle{apalike}
\bibliography{calBib}

\newpage

\section*{Tables and Figures}

\newpage

\newpage

\newpage

\begin{landscape}
\begin{table}
\tiny
\begin{tabular}{ccccccccccccccccc}
\hline
\multirow{2}{*}{$n$} & \multirow{2}{*}{$m$} & \multirow{2}{*}{$\omega^2$} & \multirow{2}{*}{$\tau^2$} & \multirow{2}{*}{GPS} & \multirow{2}{*}{Outcome} & \multirow{2}{*}{EPE} &  & \multicolumn{4}{c}{Relative Bias} &  & \multicolumn{4}{c}{95\% Coverage Probability} \\ \cline{9-12} \cline{14-17} 
 &  &  &  &  &  &  &  & No Correction & \begin{tabular}[c]{@{}c@{}}Regression\\ Calibration\end{tabular} & \begin{tabular}[c]{@{}c@{}}BART Multiple\\ Imputation\end{tabular} & \begin{tabular}[c]{@{}c@{}}GLM Multiple\\ Imputation\end{tabular} &  & No Correction & \begin{tabular}[c]{@{}c@{}}Regression\\ Calibration\end{tabular} & \begin{tabular}[c]{@{}c@{}}BART Multiple \\ Imputation\end{tabular} & \begin{tabular}[c]{@{}c@{}}GLM Multiple\\ Imputation\end{tabular} \\ \cline{1-12} \cline{14-17} 
400 & 2{,}000 & 1 & 1 &  &  &  &  & 0.35 (-0.09) & 0.28 (-0.09) & 0.21 (-0.06) & -0.03 (-0.02) &  & 0.44 (0.52) & 0.53 (0.54) & 0.75 (0.78) & 0.83 (0.90) \\
400 & 2{,}000 & 1 & 2 &  &  &  &  & 0.42 (-0.11) & 0.30 (-0.09) & 0.28 (-0.06) & -0.05 (-0.02) &  & 0.39 (0.40) & 0.52 (0.46) & 0.74 (0.73) & 0.70 (0.91) \\
400 & 2{,}000 & 2 & 1 &  &  &  &  & 0.38 (-0.12) & 0.32 (-0.11) & 0.22 (-0.08) & -0.03 (-0.02) &  & 0.39 (0.45) & 0.45 (0.48) & 0.71 (0.70) & 0.85 (0.90) \\
400 & 2{,}000 & 2 & 2 &  &  &  &  & 0.42 (-0.14) & 0.29 (-0.13) & 0.21 (-0.08) & -0.05 (-0.02) &  & 0.36 (0.35) & 0.45 (0.44) & 0.71 (0.62) & 0.75 (0.92) \\
400 & 4{,}000 & 1 & 1 &  &  &  &  & 0.30 (-0.06) & 0.21 (-0.06) & 0.20 (-0.04) & -0.03 (-0.02) &  & 0.53 (0.66) & 0.63 (0.69) & 0.78 (0.78) & 0.80 (0.91) \\
400 & 4{,}000 & 1 & 2 &  &  &  &  & 0.31 (-0.08) & 0.18 (-0.07) & 0.21 (-0.05) & -0.06 (-0.02) &  & 0.47 (0.58) & 0.63 (0.65) & 0.78 (0.76) & 0.64 (0.92) \\
400 & 4{,}000 & 2 & 1 &  &  &  &  & 0.29 (-0.09) & 0.23 (-0.08) & 0.19 (-0.06) & -0.02 (-0.02) &  & 0.47 (0.58) & 0.55 (0.64) & 0.74 (0.72) & 0.86 (0.92) \\
400 & 4{,}000 & 2 & 2 &  &  &  &  & 0.30 (-0.11) & 0.19 (-0.10) & 0.15 (-0.07) & -0.04 (-0.02) &  & 0.42 (0.48) & 0.58 (0.52) & 0.76 (0.71) & 0.69 (0.93) \\
800 & 4{,}000 & 1 & 1 &  &  &  &  & 0.29 (-0.08) & 0.21 (-0.07) & 0.12 (-0.03) & -0.04 (0.00) &  & 0.42 (0.45) & 0.53 (0.52) & 0.76 (0.75) & 0.77 (0.96) \\
800 & 4{,}000 & 1 & 2 &  &  &  &  & 0.35 (-0.11) & 0.22 (-0.09) & 0.12 (-0.04) & -0.07 (0.00) &  & 0.37 (0.38) & 0.50 (0.44) & 0.72 (0.74) & 0.62 (0.96) \\
800 & 4{,}000 & 2 & 1 &  &  &  &  & 0.36 (-0.10) & 0.29 (-0.09) & 0.15 (-0.04) & -0.04 (0.00) &  & 0.37 (0.38) & 0.41 (0.39) & 0.73 (0.71) & 0.82 (0.96) \\
800 & 4{,}000 & 2 & 2 &  &  &  &  & 0.43 (-0.12) & 0.31 (-0.10) & 0.13 (-0.04) & -0.06 (0.00) &  & 0.35 (0.30) & 0.43 (0.38) & 0.72 (0.72) & 0.65 (0.94) \\
800 & 8{,}000 & 1 & 1 &  &  &  &  & 0.20 (-0.06) & 0.15 (-0.05) & 0.10 (-0.03) & -0.04 (-0.01) &  & 0.50 (0.67) & 0.65 (0.69) & 0.78 (0.83) & 0.74 (0.94) \\
800 & 8{,}000 & 1 & 2 &  &  &  &  & 0.26 (-0.06) & 0.16 (-0.05) & 0.10 (-0.03) & -0.07 (0.00) &  & 0.44 (0.56) & 0.59 (0.60) & 0.72 (0.76) & 0.59 (0.94) \\
800 & 8{,}000 & 2 & 1 &  &  &  &  & 0.22 (-0.08) & 0.17 (-0.07) & 0.10 (-0.03) & -0.04 (-0.01) &  & 0.45 (0.52) & 0.55 (0.58) & 0.78 (0.78) & 0.78 (0.94) \\
800 & 8{,}000 & 2 & 2 &  &  &  &  & 0.28 (-0.08) & 0.19 (-0.07) & 0.11 (-0.03) & -0.07 (-0.01) &  & 0.42 (0.45) & 0.56 (0.53) & 0.73 (0.74) & 0.62 (0.94) \\ \hline
800 & 4{,}000 & 2 & 1 &  &  & \checkmark &  & 0.91 (-0.09) & 0.30 (-0.09) & 0.15 (-0.04) & -0.03 (0.00) &  & 0.30 (0.47) & 0.42 (0.40) & 0.71 (0.73) & 0.76 (0.93) \\
800 & 4{,}000 & 2 & 1 &  & \checkmark &  &  & 0.18 (-0.20) & 0.11 (-0.20) & -0.01 (-0.14) & 0.04 (0.00) &  & 0.25 (0.02) & 0.32 (0.03) & 0.55 (0.14) & 0.32 (0.96) \\
800 & 4{,}000 & 2 & 1 &  & \checkmark & \checkmark &  & 0.60 (-0.22) & 0.11 (-0.20) & -0.01 (-0.14) & 0.05 (0.00) &  & 0.19 (0.02) & 0.30 (0.02) & 0.55 (0.15) & 0.32 (0.97) \\
800 & 4{,}000 & 2 & 1 & \checkmark &  &  &  & 0.52 (-0.06) & 0.46 (-0.05) & 0.29 (-0.01) & -0.03 (0.00) &  & 0.45 (0.54) & 0.49 (0.58) & 0.75 (0.82) & 0.75 (0.93) \\
800 & 4{,}000 & 2 & 1 & \checkmark &  & \checkmark &  & 1.28 (0.02) & 0.59 (-0.02) & 0.40 (0.02) & -0.03 (0.00) &  & 0.34 (0.66) & 0.49 (0.61) & 0.74 (0.82) & 0.74 (0.93) \\
800 & 4{,}000 & 2 & 1 & \checkmark & \checkmark &  &  & 0.15 (-0.21) & 0.08 (-0.20) & -0.05 (-0.15) & -0.09 (-0.05) &  & 0.18 (0.01) & 0.28 (0.01) & 0.44 (0.08) & 0.37 (0.77) \\
800 & 4{,}000 & 2 & 1 & \checkmark & \checkmark & \checkmark &  & 0.52 (-0.25) & 0.02 (-0.23) & -0.09 (-0.16) & -0.09 (-0.05) &  & 0.15 (0.00) & 0.25 (0.00) & 0.41 (0.04) & 0.38 (0.72) \\ \hline
\end{tabular}
\caption{Relative bias and coverage probabilities averaged over the simulated ERF estimates when measurement error is present. The values in parentheses represent the statistics evaluated at $a = 11$. The check-marks indicates whether the corresponding model labeled in the column header is misspecified. The GLM approach refers to the multiple imputation implementation using a log-linear outcome model. The BART approach to multiple imputation uses a BART outcome model.} \label{tab:sim-rslt}
\end{table}
\end{landscape}

\newpage

\newpage

\begin{center}
{\huge Supplement to ``Estimating a Causal Exposure Response Function with a Continuous Error-Prone Exposure: A Study of Fine Particulate Matter and All-Cause Mortality"}
\end{center}

\beginsupplement

\normalsize
\section{MCMC Sampling Algorithm}\label{algorithm}

In this supplement, we describe the MCMC sampling steps and provide additional details about the multiple imputation measurement error correction method that we present in Section 3. We will assume a BART model for the outcome, and Gaussian linear models for the GPS and EPE with $\mu_A(\mathbf{X}_i, \phi_i) = \phi_i + \mathbf{X}^T_i \boldsymbol{\beta}$ and $\mu_S(\tilde{S}_{ij}, \mathbf{W}_{ij}) = \left(\tilde{S}_{ij}, \mathbf{W}^T_{ij}\right) \boldsymbol{\alpha}$. Much of this methodology reflects the demonstrations provided in Chapter 9 of \cite{carroll2006measurement}, if more information is desired.

Later on in the analysis stage we will require prior distributions for the parameter values in the likelihood model of (2). We suppose that exposure variance components follow conjugate inverse-gamma distributions, i.e. $\sigma^2 \sim \mathcal{IG}(r_{\sigma,1}, r_{\sigma,2})$, $\tau^2 \sim \mathcal{IG}(r_{\tau,1}, r_{\tau,2})$, and $\omega^2 \sim \mathcal{IG}(r_{\omega,1}, r_{\omega,2})$. Each of the mean parameters for the GPS and EPE models contained in the vectors $\boldsymbol{\alpha}$ and $\boldsymbol{\beta}$ are assigned a Gaussian prior with mean zero and variance $b_{\alpha}^2$ or $b_{\beta}^2$, respectively. 

Before we proceed, there is some additional notation that we will need to define. Let $\mathbf{I}$ denote a diagonal matrix with dimensions matching either the length of $\mathbf{X}_i$ or $(\tilde{S}_{ij}, \mathbf{W}^{T}_{ij})^{T}$, whichever is appropriate in the context that it appears. Finally, define the vectors $\mathbf{S}_i \equiv (S_{i1}, S_{i2}, \ldots, S_{iM_i})^{T}$, $\mathbf{A}^{(k)} \equiv (A_1^{(k)}, A_2^{(k)}, \ldots, A_n^{(k)})^{T}$, $\bar{\mathbf{Y}} \equiv (\bar{Y}_1, \bar{Y}_2, \ldots, \bar{Y}_n)^{T}$, and the matrix $\mathbf{X} = (\mathbf{X}_1, \mathbf{X}_2, \ldots, \mathbf{X}_n)^{T}$. 

To begin, initialize $A^{(1)}_i$, $\phi^{(1)}_i$, and $S^{(1)}_{ij}$ for all $j = 1,2,\ldots,M_i$ and $i = 1,2,\ldots,n$. We must also initialize the parameter values $\boldsymbol{\alpha}^{(1)}$, $\boldsymbol{\beta}^{(1)}$, $\sigma^{(1)}$, $\tau^{(1)}$, $\omega^{(1)}$, $\rho^{(1)}$, $\nu^{(1)}$, $\mathcal{F}^{(1)}_1$, $\mathcal{G}^{(1)}_1$, and $\psi^{(1)}$. For $k = 1,2,\ldots, K$ do:

\begin{itemize}
    \item \textbf{Imputation Stage}
    \begin{enumerate}
        \item Sample $S^{(k)}_{ij} \sim \mathcal{N}\left[\eta^{(k - 1)}\left(A^{(k - 1)}_i, \tilde{S}_{ij}, \mathbf{W}_{ij}\right), \left(\Sigma^{(k - 1)}_{ij}\right)^2\right]$ where \[\begin{split}
            \Sigma^{(k - 1)}_{ij} &= \left[\left(\omega^{(k - 1)}\right)^{-2} + \left(\tau^{(k - 1)}\right)^{-2}\right]^{-1/2} \quad \text{and}\\
            \eta^{(k - 1)}\left(A^{(k - 1)}_i, \tilde{S}_{ij}, \mathbf{W}_{ij}\right) &= \left(\Sigma^{(k - 1)}_{ij}\right)^2 \times \left\{\frac{A^{(k - 1)}_i}{\left(\omega^{(k - 1)}\right)^2} + \frac{\mu^{(k - 1)}_S\left(\tilde{S}_{ij}, \mathbf{W}_{ij}\right)}{\left(\tau^{(k - 1)}\right)^2}\right\}.
        \end{split}\]
        Replace known quantities of $S^{(k)}_{ij}$ when $(i,j) \in \mathcal{S}$;
        \item To sample $A^{(k)}_{i}$, we will need to consider the full conditional likelihood:
        \[ p_{A}\left(A_i\middle|\mathbf{S}_i, \mathbf{X}_i, \bar{Y}_i,\phi^{(k - 1)}_i\right) \propto p_{\bar{Y}}(\bar{Y}_i|A_i,\mathbf{X}_i) \times \exp\left\{-\frac{1}{2}\left[\frac{A_i - \mu_{A}\left(\mathbf{X}_i, \phi^{(k - 1)}_i\right)}{\sigma}\right]^2\right\} \times \exp\left[-\frac{1}{2}\sum_{j = 1}^{M_i}\left(\frac{S_{ij} - A_i}{\omega}\right)^2\right].\]
        For all $i = 1,2,\ldots,n$, sample $R_i \sim \mathcal{U}(0,1)$ and $A^{+}_i \sim \mathcal{N}(A^{(k - 1)}_{i}, \Delta^2)$ for some pre-specified value $\Delta$ that will achieve a sufficiently high acceptance probability. Set \[ A^{(k)}_{i} = \begin{cases} A^{+}_i, R_i \le \min\left(1, \delta^{(k)}_i\right) \\ A^{(k - 1)}_i, R_i > \min\left(1, \delta^{(k)}_i\right)\end{cases} \] where
        \[ \delta^{(k)}_i = \frac{p^{(k - 1)}_{\bar{Y}}\left(\bar{Y}_i\middle|A^{+}_i,\mathbf{X}_i\right) \times \exp\left\{-\frac{1}{2}\left[\frac{A^{+}_i - \mu^{(k - 1)}_{A}\left(\mathbf{X}_i,\phi^{(k - 1)}_i\right)}{\sigma^{(k - 1)}}\right]^2\right\} \times \exp\left[-\frac{1}{2} \sum_{j = 1}^{M_i}\left(\frac{S^{(k)}_{ij} - A^{+}_i} {\omega^{(k - 1)}}\right)^2\right]}{p^{(k - 1)}_{\bar{Y}}\left(\bar{Y}_i\middle|A^{(k - 1)}_i,\mathbf{X}_i\right) \times \exp\left\{-\frac{1}{2}\left[\frac{A^{(k - 1)}_i - \mu^{(k - 1)}_{A}\left(\mathbf{X}_i,\phi^{(k - 1)}_i\right)}{\sigma^{(k - 1)}}\right]^2\right\} \times \exp\left[-\frac{1}{2} \sum_{j = 1}^{M_i}\left(\frac{S^{(k)}_{ij} - A^{(k - 1)}_i} {\omega^{(k - 1)}}\right)^2\right]};\]
    \end{enumerate}
    \item \textbf{Analysis Stage}
    \begin{enumerate}[resume]
        \item Using the subset of individuals from $(i,j) \in \mathcal{S}$, sample the prediction error mean parameters 
        \[\begin{split}
            \boldsymbol{\alpha}^{(k)} \sim \mathcal{N}\vast(\left\{\sum_{(i,j)\in\mathcal{S}}\left[\left(\tilde{S}_{ij}, \mathbf{W}^T_{ij}\right)^T\left(\tilde{S}_{ij}, \mathbf{W}^T_{ij}\right)\right] + \left[\left(\tau^{(k-1)}\right)^2\middle/b^2_{\alpha}\right]\mathbf{I}\right\}^{-1}\times\sum_{(i,j) \in \mathcal{S}}\left[ S_{ij}\left(\tilde{S}_{ij}, \mathbf{W}^T_{ij}\right)^T \right], \\ \left(\tau^{(k-1)}\right)^2\left[\sum_{(i,j)\in\mathcal{S}} \left[\left(\tilde{S}_{ij}, \mathbf{W}^T_{ij}\right)^{T}\left(\tilde{S}_{ij}, \mathbf{W}^T_{ij}\right)\right] + \left[\left(\tau^{(k-1)}\right)^2\middle/b^2_{\alpha}\right]\mathbf{I}\right\}^{-1}\vast).
        \end{split} \]
        and set $\mu^{(k)}_S\left(\tilde{S}_{ij}, \mathbf{W}_{ij}\right) = \left(\tilde{S}_{ij}, \mathbf{W}^T_{ij}\right) \boldsymbol{\alpha}^{(k)}$;
        \item Sample the GPS mean parameters
         \[ \phi^{(k)}_i \sim \mathcal{N}\left(\frac{\rho^{(k - 1)} \sum_{i' = 1}^n V_{ii'}\phi_{i'}^{(k - 1)}}{\rho^{(k - 1)} \sum_{i' = 1}^n V_{ii'} + 1 - \rho^{(k - 1)}}, \frac{\left(\nu^{(k - 1)}\right)^2}{\rho^{(k - 1)} \sum_{i' = 1}^n V_{ii'} + 1 - \rho^{(k - 1)}}\right), \]
         where $V_{ii'}$ is the $i$th row and $i'$th column of $\mathbf{V}$, and
        \[\begin{split}
            \boldsymbol{\beta}^{(k)} \sim \mathcal{N}\vast(\left\{\sum_{i = 1}^n \left(\mathbf{X}_{i}\mathbf{X}^{T}_{i} \right) + \left[\left(\sigma^{(k-1)}\right)^2\middle/b^2_{\beta}\right]\mathbf{I}\right\}^{-1}\times\sum_{i = 1}^n \left[\mathbf{X}_{i}\left(A^{(k)}_{i} - \phi^{(k)}_i\right)\right], \\ \left(\sigma^{(k-1)}\right)^2\left\{\sum_{i = 1}^n \left(\mathbf{X}_{i}\mathbf{X}^{T}_{i}\right) + \left[\left(\sigma^{(k-1)}\right)^2\middle/b^2_{\beta}\right]\mathbf{I}\right\}^{-1}\vast).
        \end{split} \]
        Then set $\mu^{(k)}_A\left(\mathbf{X}_i,\phi^{(k)}_i\right) = \phi_i^{(k)} + \mathbf{X}^T_i \boldsymbol{\beta}^{(k)}$;
        \item Sample the prediction error variance \[ \left(\tau^{(k)}\right)^2 \sim \mathcal{IG}\left\{r_{\tau,1} + \frac{\left|\mathcal{S}\right|}{2}, r_{\tau,2} + \frac{1}{2} \sum_{(i, j) \in \mathcal{S}} \left[S^{(k)}_{ij} - \mu^{(k)}_S\left(\tilde{S}_{ij}, \mathbf{W}_{ij}\right)\right]^2\right\}; \]
        \item Sample the GPS variance \[ \left(\sigma^{(k)}\right)^2 \sim \mathcal{IG}\left\{r_{\sigma,1} + \frac{n}{2}, r_{\sigma,2} + \frac{1}{2} \sum_{i = 1}^n \left[A^{(k)}_{i} - \mu^{(k)}_A\left(\mathbf{X}_i, \phi^{(k)}_i\right)\right]^2\right\}; \]
        \item Sample the classical error variance \[ \left(\omega^{(k)}\right)^2 \sim \mathcal{IG}\left[r_{\omega,1} + \frac{m}{2}, r_{\omega,2} + \frac{1}{2} \sum_{i = 1}^n \sum_{j = 1}^{M_i} \left(S^{(k)}_{ij} - A^{(k)}_i\right)^2\right]; \]
        \item Sample $\rho^{(k)}$ and $\nu^{(k)}$ according to the suggestions of \cite{leroux2000estimation} and \cite{lee2013carbayes}, treating $A^{(k)}_i$ as the outcome, $\phi^{(k)}_i$ the random effect, and $\mathbf{X}_i$ the covariates;
        \item For tree $t \in \left\lbrace 1,...,T_k\right\rbrace$:
        \begin{enumerate}
            \item Sample $\mathcal{F}_t^{(k)}$ from the likelihood $p_{\mathcal{F}}\left(\mathcal{F}_t | \mathbf{A}^{(k)}, \mathbf{X}, \bar{\mathbf{Y}}; \psi^{(k-1)}\right)$ via the Metropolis-Hastings search algorithm given by \cite{chipman1998bayesian};
            \item Sample $\mathcal{G}_t^{(k)}$ from the posterior $p_{\mathcal{G}}\left(\mathcal{G}_t | \mathbf{A}^{(k)}, \mathbf{X}, \bar{\mathbf{Y}}; \psi^{(k-1)}, \mathbf{\mathcal{F}}_t^{(k)} \right)$ using draws from independent Gaussian distributions as described in \cite{chipman2010bart}. 
        \end{enumerate}
        \item Sample $\psi^{(k)}$ using a conjugate inverse-gamma prior distribution according to \cite{chipman2010bart}, supposing $N_i$ as the weights.
    \end{enumerate}
\end{itemize}
With the posterior samples of the outcome model, $\mu^{(k)}_{\bar{Y}}\left(A^{(k)}_i, \mathbf{X}_i\right)$, and the exposure imputations $A^{(k)}_i$, we can proceed to fit the ERF using the local regression method described in Section \ref{loess}.

\section{Kernel Weighted Least Squares Regression Details}\label{loess}

In this supplement, we provide the details of the smoothing step outlined in Section 3.2. Given a bandwidth $h > 0$, the pseudo-outcomes in (4) are regressed onto fixed points $a \in \mathcal{A}$ by solving  for $\hat{\theta}^{(l)}_{h}(a) = \mathbf{c}^T_{ha}(a)\hat{\boldsymbol{\lambda}}^{(l)}_{ha}$ where 
\begin{equation}\label{beta}
    \hat{\boldsymbol{\lambda}}^{(l)}_{ha} = \argmin_{\boldsymbol{\lambda} \in \Re^2} \sum_{i = 1}^n q_{ha}\left(A^{(l)}_i\right)\left\{\xi^{(l)}\left(A^{(l)}_i, \mathbf{X}_i, \bar{Y}_i\right) - \mathbf{c}^T_{ha}\left(A^{(l)}_i\right) \boldsymbol{\lambda}\right\}^2
\end{equation}
and \[ \mathbf{c}_{ha}\left(A^{(l)}_i\right) = \left[1, h^{-1}\left(A^{(l)}_i - a\right)\right]^T. \] For a given bandwidth representing the standard deviation of a Gaussian probability density function centered around $a \in \mathcal{A}$, the kernel weights are equivalent to \[ q_{ha}\left(A^{(l)}_i\right) = \frac{N_i}{h\sqrt{2\pi}} \exp\left[-\frac{\left(A^{(l)}_i - a\right)^2}{2h^2}\right]. \]

As we mentioned already in the manuscript, the values $\hat{\theta}^{(l)}_{h}(a)$ do not form a proper posterior of $\theta(a)$. Instead, think of the $\xi^{(l)}(A^{(l)}_i, \mathbf{X}_i, Y_i)$ as one of $L$ imputations for the ERF evaluated at $A^{(l)}_i$, meaning we must use combining rules intended for multiple imputation to estimate the variance of $\bar{\theta}_h(a)$. For each $l = 1,2,\ldots,L$, we require the pointwise variance estimates for each $\hat{\theta}_h^{(l)}(a)$. For this we can use a subtle modification to the locally-weighted least-squares variance estimator following the form \[ \hat{\boldsymbol{\Omega}}^{(l)}_{h}(a) = \sum_{i = 1}^n \left(\mathbf{D}^{(l)}_{ha}\right)^{-1} \left[\hat{\boldsymbol{\zeta}}^{(l)}_{ha}\left(A^{(l)}_i, \mathbf{X}_i, \bar{Y}_i\right)\right]^{\otimes2} \left(\mathbf{D}^{(l)}_{ha}\right)^{-1}, \]
where $\mathbf{D}^{(l)}_{ha} = \sum_{i = 1}^n q_{ha}\left(A^{(l)}_i\right) \mathbf{c}_{ha}\left(A^{(l)}_i\right)\mathbf{c}^T_{ha}\left(A^{(l)}_i\right)$ and
\begin{equation}\label{var-est}
\begin{split}
   \hat{\boldsymbol{\zeta}}^{(l)}_{ha}\left(A^{(l)}_i, \mathbf{X}_i, \bar{Y}_i\right) &= q_{ha}\left(A^{(l)}_i\right)\left\{\xi^{(l)}\left(A^{(l)}_i, \mathbf{X}_i, \bar{Y}_i\right) - \mathbf{c}^T_{ha}\left(A^{(l)}_i\right) \hat{\boldsymbol{\lambda}}^{(l)}_{ha}\right\}\mathbf{c}_{ha}\left(A^{(l)}_i\right) \\ &\qquad+ \int_\mathcal{A} q_{ha}\left(t\right)\left[\mu^{(l)}_{\bar{Y}}(t, \mathbf{X}_i) - \int_{\mathcal{X}} \mu^{(l)}_{\bar{Y}}(t, \mathbf{x}) d\mathbb{P}(\mathbf{x})\right] \mathbf{c}_{ha}\left(t\right)d\mathbb{Q}(t).
\end{split}
\end{equation}
Like the distribution for $\mathbf{X}_i$, the cumulative and empirical density functions for $A_i$ are written as $Q(a)$ and $\mathbb{Q}(a)$, respectively. At this point we can obtain a variance estimate using the combining rules of \cite{rubin2004multiple} with
\[ \hat{\mathbb{V}}\left[\bar{\theta}_h(a)\right] \approx \left(1+L^{-1}\right)(L - 1)^{-1}\sum_{l = 1}^L\left[\hat{\theta}^{(l)}_h(a) - \bar{\theta}_h(a) \right]^2 + L^{-1}\sum_{l = 1}^L\left[\hat{\Omega}^{(l)}_{h,(1,1)}(a)\right]^2 \] where $\hat{\Omega}^{(l)}_{h,(1,1)}(a)$ is the (1,1) element of $\hat{\boldsymbol{\Omega}}^{(l)}_{h}(a)$. This approach is ideologically similar to \cite{antonelli2020causal} who also use combining rules after regressing posterior samples onto the exposure to estimate the ERF. 

The bandwidth $h$ can be chosen by cross-validation. This is accomplished by partitioning $\left[\xi^{(l)}_i(\cdot), A^{(l)}_i\right]$, $i = 1,2,\ldots,n$, into one of several folds to find $h > 0$ that minimizes the cross-validated mean squared error via grid search.

\section{Simulation Details}\label{sim-details}

Sections \ref{correct} and \ref{misspecify} contains a complete description of the data generating mechanisms for the simulation study design in Section 4.1. We also provide the additional simulation results when at least one of the measurement error sources is set to $0$, $\tau^2 = 0$, $\omega^2 = 0$, or both in Section \ref{correct}. Contained within this table are results that partially align with the subfigures in Figure 3 (top-row and bottom-left subfigures when $n = 800$ and $m = 4000$).

\subsection{Correct Specification}\label{correct}

For cluster $i = 1,2,\ldots,n$, let $X_{i1},X_{i2},X_{i3},X_{i4} \sim \mathcal{N}(0, 1)$, $\mathbf{X}_i = (X_{i1},X_{i2},X_{i3},X_{i4})^{T}$, which we use to generate \[ A_i \sim \mathcal{N}\left(10 + 0.5X_{i1} - 0.5X_{i2} - 0.5X_{i3} + 0.5X_{i4}, \sigma^2\right) \] with $\sigma^2 = 4$. For $j = 1,2,\ldots,M_i$, we sample $S_{ij} \sim \mathcal{N}(A_i, \omega^2)$ with $\omega^2 \in \{0, 1, 2\}$. The associated predictions for $S_{ij}$ are subsequently generated with the distribution $\tilde{S}_{ij} \sim \mathcal{N}\left(0, \tau^2\right)$ where  $\tau^2 \in \{0, 1, 2\}$. The outcome counts are generated from a Poisson distribution $Y_i \sim \mathcal{P}[\mu_{Y}(A_i, \mathbf{X}_i)]$ with 
\[ \begin{split}
    \log\left[\mu_{Y}(A_i, \mathbf{X}_i)\right] &= -3 - 0.5X_{i1} - 0.25X_{i2} + 0.25X_{i3} + 0.5X_{i4} + \log(N_i)\\ &\qquad+ 0.25(A_i - 8) - 0.75\cos\left[\pi(A_i - 6)/4\right] - 0.25(A_i - 10)\times X_{i1}.
\end{split} \] 
In addition, we vary the total number of cells, $m \in \{2000, 5000\}$ when $n = 400$ and $m \in \{4000, 8000\}$ when $n = 800$. The number of measurements in each of the $n$ clusters is distributed uniformly across $M_i \in \{1,2,3,4,6,7,8,9\}$ when $n = 400$ and $m = 2000$ or $n = 800$ and $m = 4000$. When $n = 400$ and $m = 4000$ or $n = 800$ and $m = 8000$, then the number of measurements is uniformly distributed across $M_i \in \{2,4,6,8,12,14,16,18\}$. The offsets are generated from the uniform distribution $N_i \sim \mathcal{U}(10,1{,}000)$. We fix the proportion of exposures that are within the validation set $\mathcal{S}$ (i.e. when $S_{ij}$ is ``observed") to $0.1$.

\subsection{Incorrect Specification}\label{misspecify}

We would also like to examine the potential sources of bias and error that might occur when we misspecify each of the three models in (2) - the outcome model, the generalized propensity score (GPS), and the error prone exposure model (EPE). Under correct specification, the data are generated using the the models described above in Section \ref{correct}. Using the same definitions for $\mathbf{X}_i$ as above, define $\tilde{X}_{i1} = \exp\left(X_{i1}/2\right)$, $\tilde{X}_{i2} = X_{i2}(1 + X_{i1})^{-1} + 10$, $\tilde{X}_{i3} = (X_{i1}X_{i3}/25 + 0.6)^{3}$, and $\tilde{X}_{i4} = (X_{i2} + X_{i4} + 10)^2$ \citep{kang2007demystifying}. The transformations $\tilde{\mathbf{X}}_i = (\tilde{X}_{i1}, \tilde{X}_{i2}, \tilde{X}_{i3},\tilde{X}_{i4})^{T}$ are subsequently scaled and centered to have a mean of zero and a marginal variance of one. Under a misspecified GPS scenario we generate \[ A_i \sim \mathcal{N}\left[10 + 0.5\tilde{X}_{i1} - 0.5\tilde{X}_{i2} - 0.5\tilde{X}_{i3} + 0.5\tilde{X}_{i4}, \sigma^2\right],\] yet the GPS models are fit using the original covariates $\mathbf{X}_i$. Likewise, scenarios where the outcome model is misspecified means we generate $Y_{i} \sim \mathcal{P}\left[\tilde{\mu}_Y(A_i, \mathbf{X}_i)\right]$ with 
\[\begin{split}
    \log\left[\tilde{\mu}_Y(A_i, \mathbf{X}_i)\right] &= -3 - 0.5\tilde{X}_{i1} - 0.25\tilde{X}_{i2} + 0.25\tilde{X}_{i3} + 0.5\tilde{X}_{i4} + \log(N_i) \\ &\qquad+ 0.25(A_i - 8) - 0.75\cos\left[\pi(A_i - 6)/4\right] - 0.25(A_i - 10)\times \tilde{X}_{i1}.
\end{split} \] 
However, we continue to fit the outcome models with the original (i.e. untransformed) covariates similar to the scenarios where the GPS is misspecified. Finally, when the EPE model is misspecified, then we introduce prediction bias, generating $\tilde{S}_{ij} \sim \mathcal{N}\left[ S_{ij} - 1 + 0.5W_{ij1} + 0.5W_{ij2}, \tau^2\right]$ where $W_{ij1} \sim \mathcal{N}(1,2)$ and $W_{ij2} \sim \mathcal{N}(X_{i2}, 1)$. In this scenario, the na\"ive approach will produce biased results since there is no adjustment for the added prediction bias. In the regression calibration and multiple imputation approaches, however, this bias is corrected for when addressing the prediction error, and therefore the downstream estimates of the ERF should be unbiased when the other models are well approximated. For these misspecification scenarios, we fix $\sigma^2 = 4$, $\omega^2 = 1$, $\tau^2 = 0.5$, $n = 400$, and $m = 4000$.

\section{Congeniality Simulation}\label{congeniality}

In this supplement, we run a small simulation study to demonstrate how using a GPS-based estimator has almost no influence over the accuracy of the ERF estimate while adhering to rules of congeniality \citep{meng1994multiple}. As we mentioned in Section 3.3, the requirements of congeniality seem to conflict with the principles of cutting feedback in causal models fit with Bayesian methods \citep{zigler2013model}. To demonstrate how the feedback created by congeniality can affect estimators that condition on the GPS, we will need to incorporate the GPS into our pseudo-outcome. This implementation was described by \cite{kennedy2017nonparametric}, who specify the following pseudo-outcome which we have adapted slightly to fit into the multiple imputation framework:
\begin{equation}\label{pseudo-alt}
    \tilde{\xi}^{(l)}\left(A^{(l)}_i, \mathbf{X}_i, \bar{Y}_i\right) = \frac{\left[\bar{Y}_i - \mu^{(l)}_{Y}\left(A^{(l)}_i, \mathbf{X}_i\right)\right]}{p^{(l)}_A\left(A^{(l)}_i\middle|\mathbf{X}_i,\phi_i^{(l)}\right)} \int_{\mathcal{X}} p^{(l)}_A\left(A^{(l)}_i\middle|\mathbf{x},\phi_i^{(l)}\right) d\mathbb{P}(\mathbf{x}) + \int_{\mathcal{X}} \mu^{(l)}_{Y}\left(A^{(l)}_i, \mathbf{x}\right) d\mathbb{P}(\mathbf{x})
\end{equation}
The first part of this pseudo-outcome, on the left-hand side of the addition symbol, characterizes an estimator of residual error conditioned on the stabilized inverse probability weights constructed from the posterior samples of the GPS. If the outcome model is correctly specified, then the average of these debiased residuals will approach zero as $n$ approaches infinity. The second part of (\ref{pseudo-alt}), on the right-hand side of the addition symbol, is the marginalized outcome model. When the outcome model is misspecified, then the estimator of the residual error will counteract the asymptotic bias generated by the marginalized outcome model, thus allowing for an alternative means to debias the estimate so long as the GPS is correctly specified. Thus, this pseudo-outcome facilitates doubly-robust estimation when used in tandem with locally-weighted regression methods, at least in cases where imputing the exposure is not necessary.

Consider the simulation scenarios in Section \ref{sim-details}, and in particular the misspecification scenarios in Section \ref{misspecify}. We will once again fit the model in (2), but substitute $\xi^{(l)}(\cdot)$ with $\tilde{\xi}^{(l)}(\cdot)$. Moreover, in an effort to better examine the doubly-robust properties of the pseudo-outcome in (\ref{pseudo-alt}), which can be masked by the flexibility of a BART model, we will fit the likelihood in (2) assuming $p_{\bar{Y}}(\bar{Y}_i|A_i, \mathbf{X}_i)$ is a Poisson model, $Y_i \sim \mathcal{P}\left[\mu_Y(A_i, \mathbf{X}_i)\right]$, where 
\begin{equation}\label{glm}
    \log[\mu_Y(A_i, \mathbf{X}_i)] = \gamma_0 + \gamma_1(A_i - 10) + \gamma_2\cos\left[\pi(A_i - 6)/4\right] + \gamma_3(A_i - 10)X_{i1} + \boldsymbol{\gamma}_4^T\mathbf{X}_i + \log(N_i). 
\end{equation} 
Note that the simulated outcome data without misspecification is generated using the same model, where $\gamma_0$--$\boldsymbol{\gamma}_4$ are known. Otherwise, the smoothing steps outlined in Section \ref{loess} remain unchanged other than to substitute $\mu^{(l)}_{\bar{Y}}\left(A^{(l)}_i, \mathbf{X}_i\right)$ with the form in (\ref{glm}). The Gaussian models that we assumed for $p_A(A_i|\mathbf{X}_i)$ with $\phi_i = 0$ for all $i = 1,2,\ldots,n$, $p_S(S_{ij}|A_i)$, and $p_S(S_{ij}|\tilde{S}_{ij}, \mathbf{W}_{ij})$ in Section 4 are also implemented in the same fashion here. The algorithm described in Section \ref{algorithm} is used to generate posterior draws of the latent exposures and unknown parameter values.

\begin{figure}[H]
	\centering
	\includegraphics[scale = 0.5]{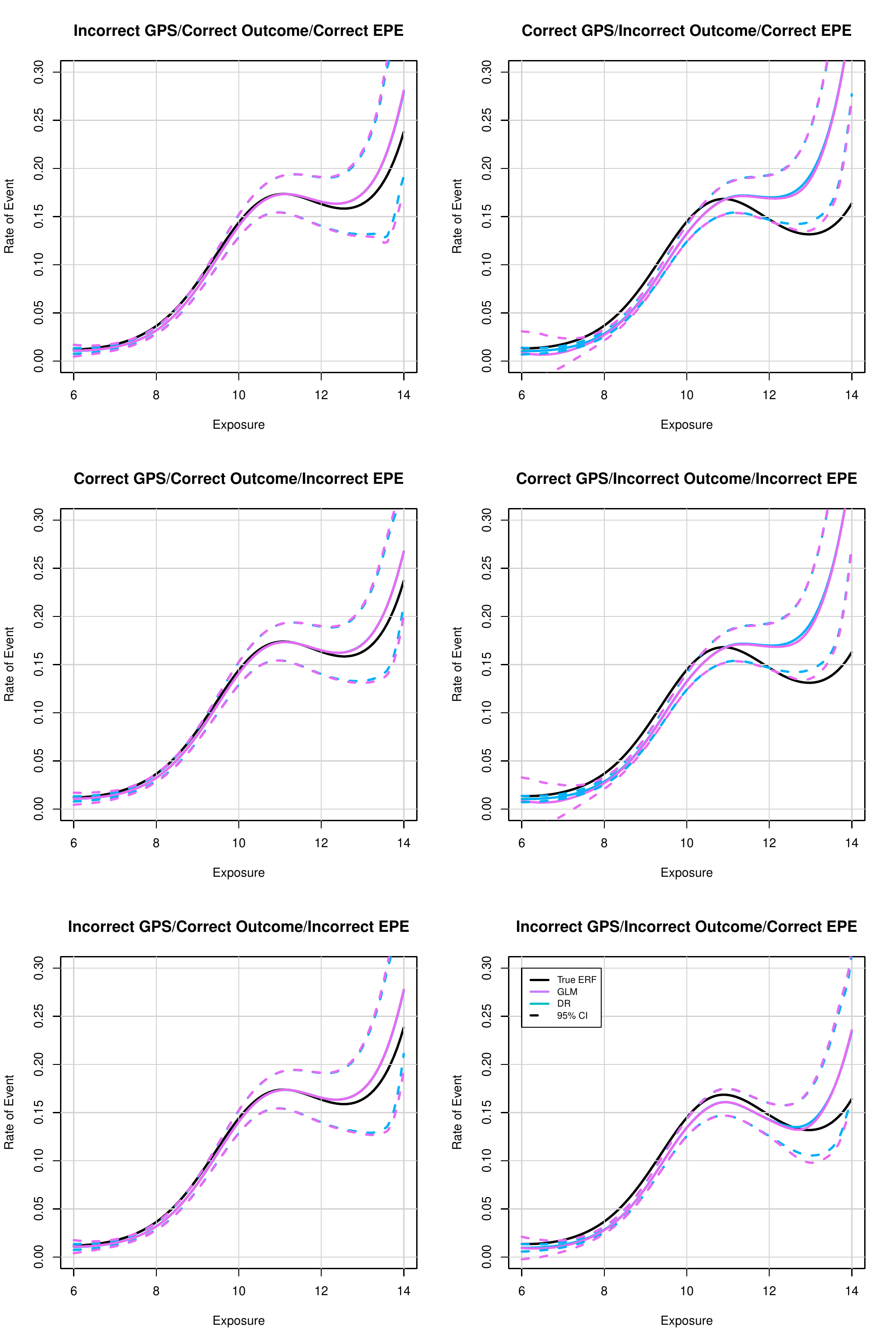}
	\caption{ERF estimates averaged over $500$ simulated iterations when $n = 800$ and $M = 4000$ under various model misspecification conditions.}\label{fig:congenial}
\end{figure}

We are most interested in the scenarios where the outcome model is misspecified and the GPS is correctly specified. For an estimator to be doubly-robust, it must produce unbiased estimates in these scenarios. There are two curves for every subfigure in Figure \ref{fig:congenial}, one where we use $\xi^{(l)}(\cdot)$ and the other where we use $\tilde{\xi}^{(l)}(\cdot)$. Both curves use Gaussian kernel weighted least squares regression to project the respective pseudo-outcomes onto $a \in \mathcal{A}$. Examining the scenarios where the GPS is correctly specified but the outcome model is misspecified, we can see that the two curves are nearly identical despite one using a doubly-robust form. Therefore, the utility of a doubly-robust estimator in this framework yields little to no benefit. The observed bias that we detect when the outcome model is misspecified and the GPS model is correctly specified, we contend, is due to the feedback created between the GPS and the outcome model from the requirements of congeniality to support multiple imputation.

The above analysis guided the development of our framework in the main manuscript. In an effort to limit any potential deleterious impacts created by feedback that may or may not exist given our limited testing, we decided to refrain from incorporating the GPS into our estimator of the ERF, even though we lose the possibility of achieving the doubly-robust property. However, from this simulation study, we did not see any benefit to using a doubly-robust estimator since it would seem that the outcome model needs to be correctly specified anyway. To make up for this loss, we decided to incorporate BART into our estimator of the ERF. BART is a nonparametric approach that provides added flexibility in cases of outcome model misspecification. We concede that more work is required to better understand how to ameliorate the feedback issue for GPS-based estimators of the ERF with a latent exposure variable.

\newpage

\section*{Tables and Figures}

\begin{landscape}
\begin{table}[]
\scriptsize
\begin{tabular}{cccccccccccccc}
\hline
\multirow{2}{*}{$n$} & \multirow{2}{*}{$m$} & \multirow{2}{*}{$\omega^2$} & \multirow{2}{*}{$\tau^2$} &  & \multicolumn{4}{c}{Relative Bias} &  & \multicolumn{4}{c}{95\% Coverage Probability} \\ \cline{6-9} \cline{11-14} 
 &  &  &  &  & No Correction & \begin{tabular}[c]{@{}c@{}}Regression\\ Calibration\end{tabular} & \begin{tabular}[c]{@{}c@{}}BART Multiple\\ Imputation\end{tabular} & \begin{tabular}[c]{@{}c@{}}GLM Multiple\\ Imputation\end{tabular} &  & No Correction & \begin{tabular}[c]{@{}c@{}}Regression\\ Calibration\end{tabular} & \begin{tabular}[c]{@{}c@{}}BART Multiple \\ Imputation\end{tabular} & \begin{tabular}[c]{@{}c@{}}GLM Multiple\\ Imputation\end{tabular} \\ \cline{1-9} \cline{11-14} 
400 & 4,000 & 0 & 0 &  & 0.19 (-0.03) & 0.20 (-0.03) & 0.20 (-0.03) & 0.02 (-0.02) &  & 0.65 (0.75) & 0.65 (0.72) & 0.75 (0.80) & 0.90 (0.87) \\
400 & 4,000 & 0 & 1 &  & 0.31 (-0.04) & 0.23 (-0.03) & 0.25 (-0.02) & 0.00 (-0.02) &  & 0.56 (0.64) & 0.66 (0.67) & 0.79 (0.78) & 0.91 (0.88) \\
400 & 4,000 & 0 & 2 &  & 0.26 (-0.07) & 0.14 (-0.07) & 0.15 (-0.05) & 0.00 (-0.02) &  & 0.51 (0.64) & 0.63 (0.69) & 0.80 (0.78) & 0.90 (0.88) \\
400 & 4,000 & 1 & 0 &  & 0.20 (-0.06) & 0.21 (-0.06) & 0.18 (-0.05) & 0.03 (-0.02) &  & 0.57 (0.70) & 0.57 (0.70) & 0.74 (0.76) & 0.93 (0.92) \\
400 & 4,000 & 2 & 0 &  & 0.24 (-0.07) & 0.24 (-0.08) & 0.20 (-0.06) & 0.02 (-0.02) &  & 0.50 (0.62) & 0.50 (0.62) & 0.69 (0.75) & 0.93 (0.90) \\
400 & 2,000 & 0 & 0 &  & 0.34 (0.01) & 0.35 (0.01) & 0.39 (0.02) & 0.02 (-0.02) &  & 0.64 (0.75) & 0.64 (0.76) & 0.74 (0.78) & 0.91 (0.90) \\
400 & 2,000 & 0 & 1 &  & 0.27 (-0.06) & 0.17 (-0.06) & 0.19 (-0.04) & 0.00 (-0.02) &  & 0.53 (0.66) & 0.64 (0.66) & 0.76 (0.75) & 0.88 (0.92) \\
400 & 2,000 & 0 & 2 &  & 0.31 (-0.10) & 0.18 (-0.08) & 0.19 (-0.05) & -0.01 (-0.02) &  & 0.42 (0.48) & 0.59 (0.54) & 0.76 (0.71) & 0.85 (0.86) \\
400 & 2,000 & 1 & 0 &  & 0.25 (-0.08) & 0.24 (-0.07) & 0.19 (-0.06) & 0.01 (-0.02) &  & 0.49 (0.58) & 0.49 (0.62) & 0.69 (0.73) & 0.93 (0.90) \\
400 & 2,000 & 2 & 0 &  & 0.35 (-0.09) & 0.35 (-0.09) & 0.24 (-0.07) & 0.02 (-0.02) &  & 0.44 (0.47) & 0.45 (0.52) & 0.67 (0.72) & 0.92 (0.89) \\
800 & 8,000 & 0 & 0 &  & 0.09 (-0.02) & 0.10 (-0.02) & 0.10 (-0.02) & 0.00 (-0.01) &  & 0.69 (0.81) & 0.68 (0.80) & 0.78 (0.84) & 0.93 (0.94) \\
800 & 8,000 & 0 & 1 &  & 0.13 (-0.04) & 0.08 (-0.03) & 0.10 (-0.02) & -0.01 (-0.01) &  & 0.57 (0.76) & 0.68 (0.78) & 0.79 (0.81) & 0.88 (0.94) \\
800 & 8,000 & 0 & 2 &  & 0.18 (-0.06) & 0.10 (-0.05) & 0.10 (-0.02) & -0.01 (-0.01) &  & 0.49 (0.62) & 0.55 (0.70) & 0.78 (0.84) & 0.87 (0.92) \\
800 & 8,000 & 1 & 0 &  & 0.16 (-0.04) & 0.15 (-0.04) & 0.12 (-0.02) & 0.01 (-0.01) &  & 0.57 (0.74) & 0.56 (0.74) & 0.75 (0.84) & 0.97 (0.96) \\
800 & 8,000 & 2 & 0 &  & 0.21 (-0.05) & 0.21 (-0.05) & 0.15 (-0.03) & 0.01 (-0.01) &  & 0.50 (0.62) & 0.52 (0.68) & 0.72 (0.80) & 0.97 (0.96) \\
800 & 4,000 & 0 & 0 &  & 0.11 (-0.02) & 0.11 (-0.01) & 0.12 (-0.02) & 0.01 (0.00) &  & 0.69 (0.82) & 0.70 (0.84) & 0.79 (0.90) & 0.94 (0.94) \\
800 & 4,000 & 0 & 1 &  & 0.21 (-0.05) & 0.14 (-0.04) & 0.13 (-0.03) & -0.02 (0.00) &  & 0.50 (0.62) & 0.65 (0.67) & 0.76 (0.76) & 0.87 (0.93) \\
800 & 4,000 & 0 & 2 &  & 0.28 (-0.09) & 0.18 (-0.07) & 0.11 (-0.03) & -0.03 (-0.01) &  & 0.39 (0.42) & 0.55 (0.56) & 0.74 (0.76) & 0.76 (0.94) \\
800 & 4,000 & 1 & 0 &  & 0.19 (-0.06) & 0.18 (-0.06) & 0.13 (-0.04) & 0.01 (-0.01) &  & 0.48 (0.57) & 0.48 (0.56) & 0.72 (0.76) & 0.96 (0.94) \\
800 & 4,000 & 2 & 0 &  & 0.28 (-0.09) & 0.28 (-0.08) & 0.17 (-0.04) & 0.01 (-0.01) &  & 0.42 (0.45) & 0.42 (0.48) & 0.69 (0.76) & 0.96 (0.94) \\ \hline
\end{tabular}
\caption{Relative bias and coverage probabilities averaged over the simulated ERF estimates when either $\tau^2 = 0$, $\omega^2 = 0$, or both. The values in parentheses represent these same statistics evaluated at $a = 11$.}
\end{table}
\end{landscape}

\end{document}